\documentclass[nofootinbib,aps,11pt]{revtex4-1}
\usepackage{graphicx}
\usepackage{hyperref}
\usepackage{cancel}
\usepackage{amssymb}
\usepackage{textcomp}
\usepackage{amsmath}
\usepackage{bm}
\usepackage{times}
\usepackage{epsfig}
\usepackage{color}
\usepackage{multirow}

\begin{document}
\title{\Large Novel Signatures of Heavy Neutral Lepton at Muon Collider}
\bigskip
\author{Xue-Xin Zhang}
\author{Zhi-Long Han}
\email{sps\_hanzl@ujn.edu.cn}
\author{Fei Huang}
\email{sps\_huangf@ujn.edu.cn}
\author{Honglei Li}
\email{sps\_lihl@ujn.edu.cn}
\affiliation{
	School of Physics and Technology, University of Jinan, Jinan, Shandong 250022, China
}
\date{\today}

\begin{abstract}
The Higgs-strahlung process $\ell^+\ell^-\to Z h$ is one of the most important production channels of the standard model Higgs boson $h$ at the lepton colliders. The cross section reaches the maximum value slightly above the threshold $\sqrt{s}\sim m_Z+m_h$, and decreases as $\sim 1/s$ at high energies. In the gauged extension models, the new gauge boson $Z'$ and heavy Higgs boson $H$ exist after the symmetry breaking. The heavy Higgs-strahlung process $\ell^+\ell^-\to Z' H$ would also reach the maximum cross section around the threshold $\sqrt{s}\sim m_{Z'}+m_H$. Therefore, the future high energy lepton colliders, such as the TeV scale muon collider, are promising to probe this new process. If heavy neutral lepton $N$ is introduced to generate the tiny neutrino masses via seesaw mechanism, novel signatures could arise from $\mu^+\mu^-\to Z' H \to NN +NN \to 4 \mu^\pm +4J$ and $\mu^+\mu^-\to Z' H \to \mu^+\mu^- +NN \to  \mu^\pm + \mu^\mp  +2J$, where the fat-jets $J$ come from the hadronic decay of $W$ bosons. In this paper, we investigate the same-sign tetralepton signature $4\mu^\pm+4J$ and the same-sign trilepton signature $ 3\mu^\pm + \mu^\mp  + 2J$ at the  3 TeV and 10 TeV muon collider.

\end{abstract}

\maketitle

\section{Introduction}

The observations of the tiny neutrino mass and mixing \cite{Super-Kamiokande:1998kpq,SNO:2002tuh,DayaBay:2012fng} are the direct evidence of physics beyond the standard model. Sub-eV scale neutrino mass can be naturally generated via the seesaw mechanism by introducing heavy neutral leptons $N$ \cite{Minkowski:1977sc,Mohapatra:1979ia,Schechter:1980gr,Schechter:1981cv}. The promising lepton number violation signature, such as $pp\to \ell^\pm N\to \ell^\pm \ell^\pm jj$ \cite{Han:2006ip}, is extensively studied in previous researches \cite{Deppisch:2015qwa,Cai:2017mow}. Being singlets under the standard model symmetry, the production of heavy neutral leptons at colliders is governed by the mixing parameter with the light neutrinos $V_{\ell N}$ \cite{ATLAS:2019kpx,ATLAS:2022atq,CMS:2022fut,CMS:2023jqi,CMS:2024xdq}. However, the canonical seesaw prediction $|V_{\ell N}|^2\simeq m_\nu/m_N$ is heavily suppressed by the tiny neutrino mass, which can be hardly tested at colliders \cite{Abdullahi:2022jlv}.

This motivates the gauge extension of the seesaw mechanism, such as the left-right symmetric model \cite{Pati:1974yy,Mohapatra:1974gc,Senjanovic:1975rk} and the gauged $U(1)_{B-L}$ model \cite{Davidson:1978pm,Marshak:1979fm,Mohapatra:1980qe}. Charged under the new gauge symmetry, the heavy neutral leptons have new interactions mediated by the new gauge bosons and new Higgs bosons, which could induce the lepton number violation signature at colliders that is independent of the mixing parameter $V_{\ell N}$ \cite{Fargion:1995qb,Fargion:1999ss,Huitu:2008gf,Basso:2008iv,Han:2012vk,Maiezza:2015lza,Nemevsek:2018bbt}. However, with direct couplings to quarks, the new gauge bosons in the left-right symmetric and gauged $U(1)_{B-L}$ models are tightly constrained by current LHC searches \cite{ATLAS:2019erb,ATLAS:2023cjo,CMS:2021dzb,CMS:2023ooo}. An alternative pathway is the leptophilic gauge symmetry, such as the $U(1)_{L_\mu-L_\tau}$ symmetry \cite{He:1990pn,Foot:1994vd}. At LHC, the new gauge boson $Z'$ can be produced by final state radiation from muon or tau flavor leptons, so the current searches are only sensitive to $m_{Z'}$ less than 81~GeV \cite{ATLAS:2023vxg,ATLAS:2024uvu}. In the future, the proposed multi-TeV muon collider is promising to test the gauged $U(1)_{L_\mu-L_\tau}$ symmetry \cite{Huang:2021nkl,Sun:2023ylp,Dasgupta:2023zrh}.

Various signatures of heavy neutral lepton can be produced at the muon collider. With relatively large mixing parameter $|V_{\ell N}|^2>10^{-6}$, the promising signature is $\mu^+\mu^-\to N\nu \to \ell jj \nu$ \cite{Chakraborty:2022pcc,Mekala:2023diu,Kwok:2023dck,Li:2023tbx,Wang:2023zhh,Cao:2024rzb}. Similar signature could also be induced by the dipole operator $\bar{\nu}\sigma^{\mu \nu} N F_{\mu \nu}$ \cite{Barducci:2024kig,Frigerio:2024jlh,Brdar:2025iua,Vignaroli:2025pwn}. Lepton number violation signature can originate from the vector boson scattering process $W^\pm Z/\gamma \to \ell^\pm N\to \ell^\pm \ell^\pm jj$ \cite{Li:2023lkl,Mikulenko:2023ezx,Dehghani:2025xkd}, and from the associated production process $\mu^+\mu^-\to NW^\pm\ell^\mp\to W^\pm W^\pm \ell^\mp\ell^\mp$ \cite{Antonov:2023otp}.  At the same sign muon collider, heavy neutral lepton will contribute to the $t$-channel process $\mu^+\mu^+\to W^+W^+$ \cite{Jiang:2023mte,Das:2024kyk,deLima:2024ohf,Bhattacharya:2025xwv}. Meanwhile, lepton number violation signature from heavy neutral lepton pair production $NN\to \ell^\pm \ell^\pm 4j$ can be mediated by the new gauge boson $Z'$ \cite{Liu:2021akf,He:2024dwh,Bi:2024pkk}, the axion-like particle $a$ \cite{deGiorgi:2022oks,Marcos:2024yfm}, and  the Higgs bosons $H^\pm/h/H$ \cite{Batra:2023ssq,Yang:2025jxc,Das:2025rlt,Yan:2026wrb,Yang:2026jde}, which is not suppressed by the seesaw induced mixing parameter.

It is well known that the Higgs-strahlung process $\ell^+\ell^-\to Zh$ is the most important production channel of the standard model Higgs $h$ at the lepton colliders \cite{Kilian:1995tr,Carena:2002es}. The cross section reaches the maximum value slightly above the threshold $\sqrt{s}\sim m_Z+m_h$, and decreases as $1/s$ at high energies. Therefore, the dominant production channel of Higgs $h$ becomes the vector boson fusion process $\mu^+\mu^-\to \nu_\mu \bar{\nu}_\mu h$ at the TeV scale muon collider \cite{Andreetto:2024rra}. In the gauged $U(1)$ extension of seesaw models, the new gauge boson $Z'$ and heavy Higgs $H$ exist after the symmetry breaking. The new Higgs-strahlung process $\mu^+\mu^-\to Z'H$ would reach the maximum cross section around the threshold $\sqrt{s}\sim m_{Z'}+m_H$ \cite{Basso:2010si}. Hence, the multi-TeV scale muon collider is quite promising to probe new $Z'$ and $H$ above the electroweak scale \cite{Das:2022mmh}. Cascade decays of the new particles could generate novel signatures as $\mu^+\mu^-\to Z'H\to NN+NN\to 4\mu^\pm +4J$ and $\mu^+\mu^-\to Z' H \to \mu^+\mu^- +NN \to  3\mu^\pm + \mu^\mp  +2J$, where the fat-jets $J$ come from the hadronic decay of $W$ bosons. In this paper, we investigate the same-sign tetralepton signature $4\mu^\pm+4J$ and the same-sign trilepton signature $ 3\mu^\pm + \mu^\mp + 2J$ at the  3 TeV and 10 TeV muon collider in the less constrained gauged $U(1)_{L_\mu-L_\tau}$ symmetry.

This paper is organized as follows. In Section \ref{SEC:MD}, we review the gauged $U(1)_{L_\mu-L_\tau}$ symmetry extended seesaw model. The decay properties of the new gauge boson $Z'$ and heavy Higgs $H$ are studied in Section \ref{SEC:DC}. The cross section of the heavy Higgs-strahlung at muon collider $\mu^+\mu^-\to Z'H$ is calculated in Section \ref{SEC:CS}. The novel same-sign tetralepton signature $4\mu^\pm+4J$ and same-sign trilepton $ 3\mu^\pm + \mu^\mp  + 2J$ are investigated in Section \ref{SEC:SG1} and Section \ref{SEC:SG2}, respectively. Finally, the conclusion is in Section \ref{SEC:CL}.

\section{The Model}\label{SEC:MD}

Among various gauged extensions of the standard model \cite{KA:2023dyz,Nomura:2024pwr}, the $U(1)_{L_\mu-L_\tau}$ symmetry is less constrained at present due to the lack of direct couplings of $Z'$ to quarks and electron. While the muon collider is promising to test this $U(1)_{L_\mu-L_\tau}$ symmetry. The gauged $U(1)_{L_\mu-L_\tau}$ seesaw model contains three heavy neutral leptons $(N_e,N_\mu,N_\tau)$. A scalar singlet $S$ with $U(1)_{L_\mu-L_\tau}$ charge $+1$ breaks the symmetry spontaneously. The new kinetic terms are \cite{Arcadi:2018tly,Hapitas:2021ilr}
\begin{equation}
	\mathcal{L}\supset-\frac{1}{4}F'_{\mu \nu}F'^{\mu\nu}-\frac{\epsilon}{2}F'_{\mu\nu}F^{\mu\nu} + |D_\mu S|^2,
\end{equation}
where $F'_{\mu\nu}$ and $F_{\mu\nu}$ are the field strength of the $U(1)_{L_\mu-L_\tau}$ and $U(1)_Y$ symmetry. $D_\mu=\partial_\mu-ig'Z'_\mu$ is the covariant derivative with $g'$ being the $L_\mu-L_\tau$ gauge coupling constant. The second term induces the mixing between the standard model gauge boson $Z$ and new gauge boson $Z'$ \cite{Yin:2021rlr,Lu:2023jlr}. For simplicity, we assume vanishing kinetic mixing $\epsilon=0$ in the following studies. After $S$ obtains nonzero vacuum expectation value $v_S$,  $Z'$ boson develops mass as \cite{Nomura:2020vnk}
\begin{equation}
	m_{Z'}=g' v_S.
\end{equation}

The interactions of the new boson $Z'$ with fermions are
\begin{equation}
	\mathcal{L}\supset g' (\bar{\mu}\gamma^\mu\mu-\bar{\tau}\gamma^\mu \tau +\bar{\nu}_\mu \gamma^\mu P_L\nu_\mu-\bar{\nu}_\tau \gamma^\mu P_L\nu_\tau+\bar{N}_\mu \gamma^\mu P_R N_\mu-\bar{N}_\tau \gamma^\mu P_R N_\tau)Z'_\mu.
\end{equation}

The scalar potential of the Higgs doublet $\Phi$ and singlet $S$ under the $U(1)_{L_\mu-L_\tau}$ symmetry is \cite{Das:2022mmh}
\begin{equation}
	V=-\mu_\Phi^2 \Phi^\dag \Phi-\mu_S^2 S^*S+\lambda_\Phi (\Phi^\dag\Phi)^2+\lambda_S (S^*S)^2+\lambda_{\Phi S} (\Phi^\dag\Phi)(S^*S).
\end{equation}
After the spontaneous symmetry breaking, the scalars can be denoted as
\begin{align}
 \Phi=	\left(
	\begin{array}{c}
		0 \\
		\frac{v_0+\phi^0}{\sqrt{2}}
	\end{array}\right) ,
	S=\frac{v_s+s^0}{\sqrt{2}}.
\end{align}

The squared mass matrix for the CP-even scalar bosons is
\begin{align}
	M^2(\phi^0,s^0)=\begin{pmatrix}
		2\lambda_\Phi v_0^2 & \lambda_{\Phi S}v_0v_s \\ 
		\lambda_{\Phi S}v_0v_s & 2\lambda_S v_s^2 
	\end{pmatrix},
\end{align}
which can be diagonalized by an orthogonal transformation as
\begin{align}
	\left(
	\begin{array}{c}
		h \\
		H
	\end{array}\right) = 
	\left(
	\begin{array}{c c}
		\cos\alpha & -\sin\alpha \\
		\sin\alpha & \cos\alpha
	\end{array}\right)
	\left(
	\begin{array}{c}
		\phi^0 \\
		s^0
	\end{array}
	\right),~\tan(2\alpha)=\frac{\lambda_{\Phi S}v_0v_s}{\lambda_{S}v_s^2-\lambda_{\Phi}v_0^2}.
\end{align}
The corresponding masses for the two physical Higgs states $h,H$ are
\begin{align}
	m_{h}^2=\lambda_{\Phi}v_0^2+\lambda_{S}v_s^2-\sqrt{(\lambda_{\Phi}v_0^2-\lambda_{S}v_s^2)^2+(\lambda_{\Phi S}v_0v_s)^2},\\
	m_{H}^2=\lambda_{\Phi}v_0^2+\lambda_{S}v_s^2+\sqrt{(\lambda_{\Phi}v_0^2-\lambda_{S}v_s^2)^2+(\lambda_{\Phi S}v_0v_s)^2}.
\end{align}
In this paper, we regard $h$ as the standard model Higgs. Under the current LHC constraints, $\sin\alpha\lesssim0.1$ should be satisfied for $m_H$ above the electroweak scale \cite{Lane:2024vur}. To satisfy this constraint, we assume $\sin\alpha=0$ for simplicity.

The Yukawa interactions and mass terms in the minimal scenario are given by  \cite{Asai:2018ocx}
\begin{eqnarray}
	\mathcal{L}&\supset& - y_e \bar{L}_e \tilde{\Phi} N_e - y_\mu \bar{L}_\mu \tilde{\Phi} N_\mu - y_\tau \bar{L}_\tau \tilde{\Phi} N_\tau -\frac{1}{2} M_{ee} \overline{N^c_e}N_e \\ \nonumber
	&&-y_{e\mu} S^* \overline{N^c_e}N_\mu -y_{e\tau} S \overline{N^c_e}N_\tau -M_{\mu\tau}\overline{N^c_\mu}N_\tau+ \text{h.c.},
\end{eqnarray}
where $L_e,L_\mu,L_\tau$ are the lepton doublets, and $\tilde{\Phi}=i\tau_2 \Phi^*$.
The resulting Dirac neutrino mass matrix and heavy neutral lepton mass matrix are 
\begin{align}
	M_D=\left(
	\begin{array}{c c c}
		\frac{y_e v_0}{\sqrt{2}} & 0 & 0\\
		0 & \frac{y_\mu v_0}{\sqrt{2}}&0 \\
		0 & 0 & \frac{y_\tau v_0}{\sqrt{2}}
	\end{array}
	\right), \quad
	M_N =
	\left(
	\begin{array}{c c c}
		M_{ee} & \frac{y_{e\mu}v_s}{\sqrt{2}} & \frac{y_{e\tau}v_s}{\sqrt{2}}\\
		\frac{y_{e\mu}v_s}{\sqrt{2}} & 0& M_{\mu\tau} \\
		\frac{y_{e\tau}v_s}{\sqrt{2}} & M_{\mu\tau} & 0
	\end{array}\right).
\end{align}
Light neutrino masses are generated via the type-I seesaw mechanism
\begin{equation}\label{Eq:SS}
	M_\nu\simeq - M_D M_N^{-1} M_D^T.
\end{equation}

\section{Decay Properties}\label{SEC:DC}

\begin{figure}
	\begin{center}
		\includegraphics[width=0.45\linewidth]{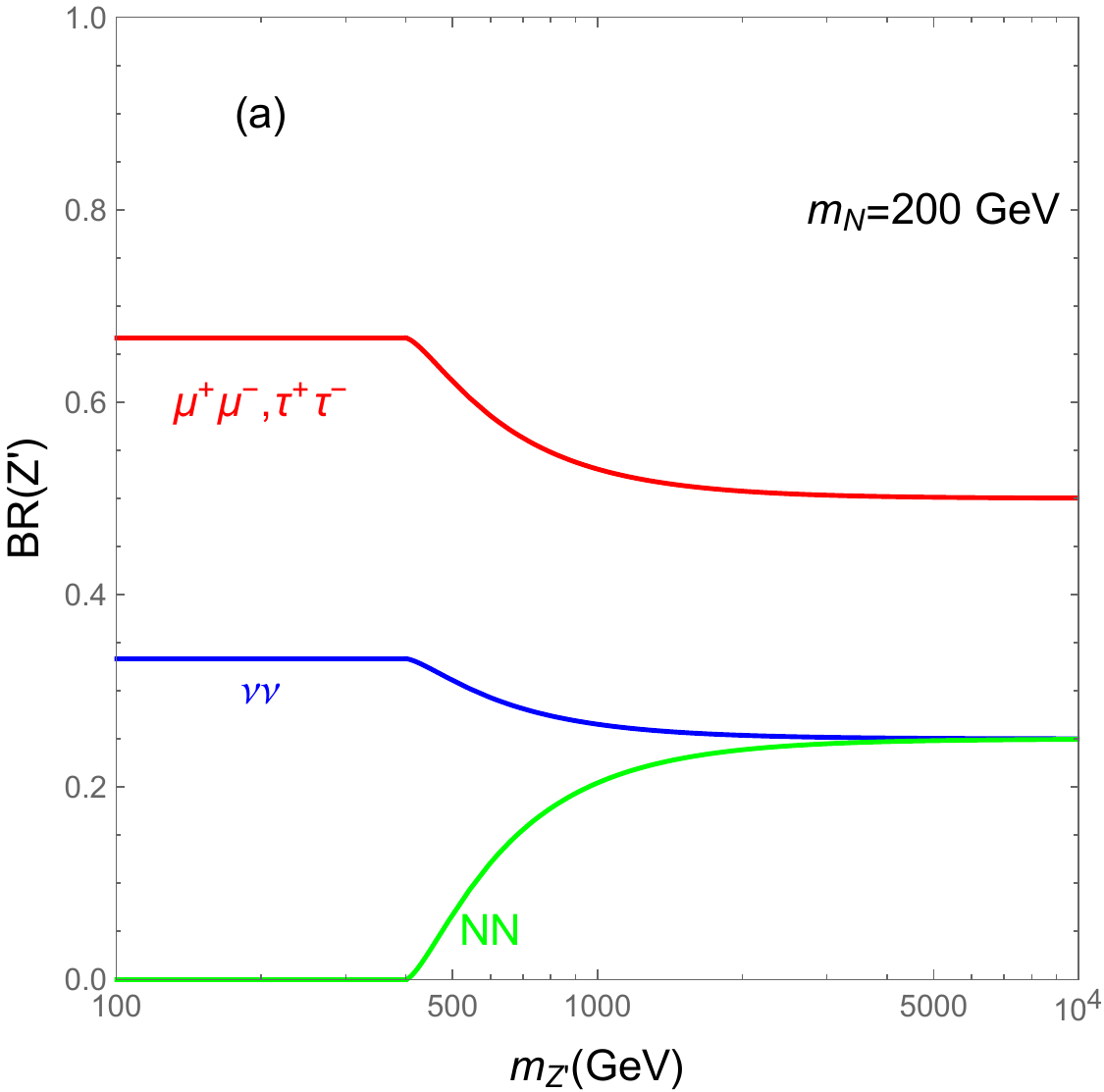}
		\includegraphics[width=0.45\linewidth]{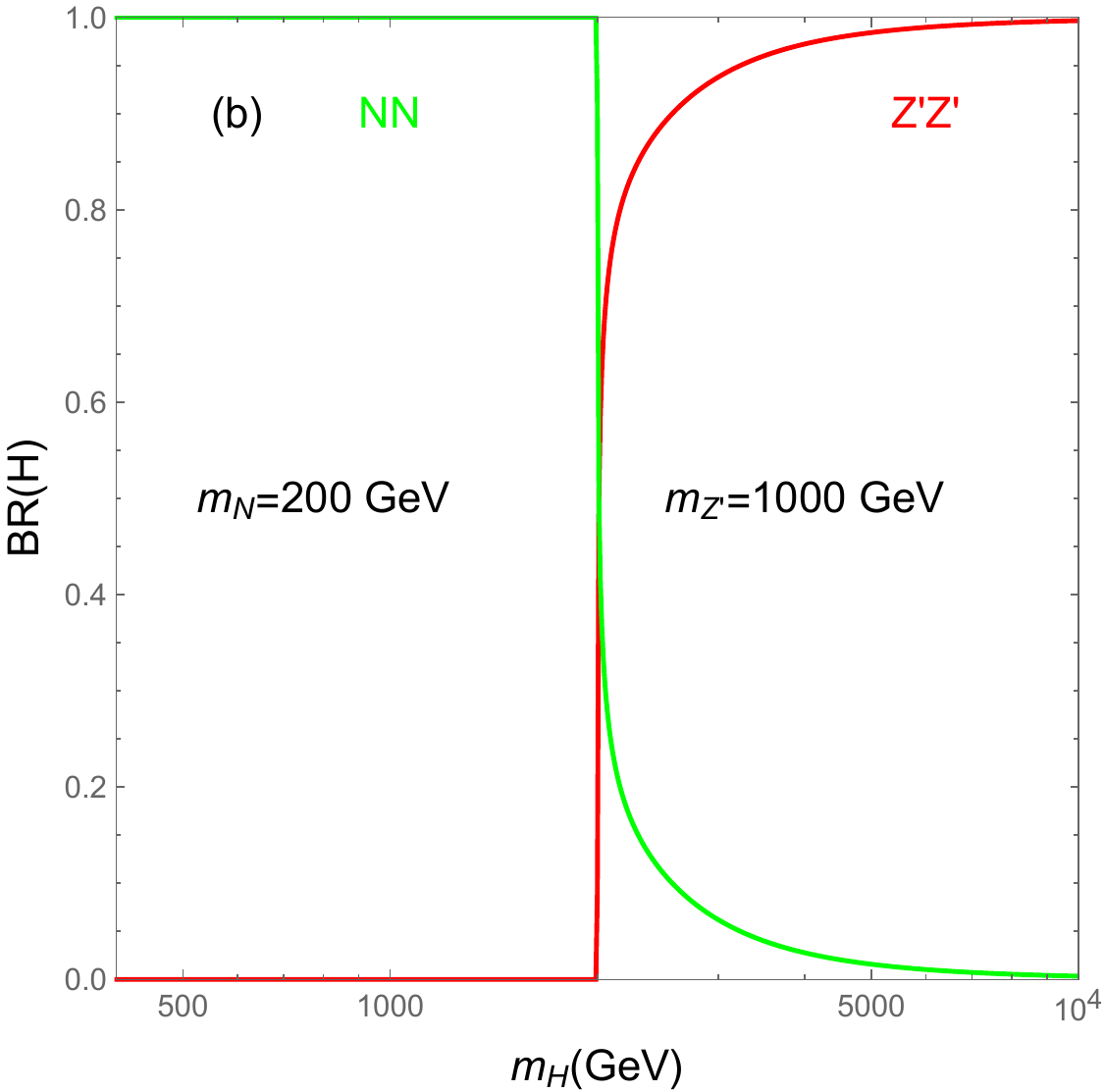}
	\end{center}
	\caption{Branching ratio of new gauge boson $Z'$ in panel (a) and heavy Higgs $H$ in panel (b). We have set $m_N=200~\rm{GeV}$ in panel (a). Meanwhile, $m_{Z'}$ is fixed to 1000 GeV in panel (b).}
	\label{fig1}
\end{figure}

In this section, we study the decay properties of the new gauge boson $Z'$ and heavy Higgs $H$.
The partial decay widths of $Z'$ are calculated as
\begin{eqnarray}
	\Gamma(Z'\to \ell^+\ell^-)&=&\frac{g'^2}{12\pi}m_{Z'}, \\
	\Gamma(Z'\to \nu_\ell \nu_\ell)&=&\frac{g'^2}{24\pi}m_{Z'},\\
	\Gamma(Z'\to N  N )&=& \frac{g'^2}{24\pi}m_{Z'}\left(1-4\frac{m_{N }^2}{m_{Z'}^2}\right).
\end{eqnarray}

In the decoupling limit $\sin\alpha=0$, there are only two viable decay modes of the heavy Higgs, i.e., $H\to NN$ and $H\to Z'Z'$ \cite{Nomura:2020vnk}. The corresponding decay widths are 
\begin{eqnarray}
	\Gamma(H\rightarrow NN)&=& \frac{m_N^2 m_H}{4\pi v_s^2} \left(1-\frac{4m_N^2}{m_H^2}\right)^{3/2},\\
	\Gamma(H\rightarrow Z'Z')&=&\frac{m_H^3 }{32\pi v_s^2} \left(1-\frac{4m_{Z'}^2}{m_H^2}+ \frac{12m_{Z'}^4}{m_H^4}\right)\left(1-\frac{4m_{Z'}^2}{m_H^2}\right)^{1/2}.
\end{eqnarray}

In Figure \ref{fig1}, we show the numerical results for the branching ratio of the new gauge boson $Z'$ and heavy Higgs $H$. Decays of $Z'$ into charged lepton pair $\mu^+\mu^-,\tau^+\tau^-$ are the dominant channel. In the limit of $m_N\ll m_{Z'}$, we approximately have BR$(Z'\to \nu \nu)\simeq\text{BR}(Z'\to NN)\simeq0.25$. For the heavy Higgs, the $H\to NN$ is the only decay mode when $m_H<2m_{Z'}$. Once kinematically allowed, the $H\to Z'Z'$ quickly becomes the dominant decay mode, as $\Gamma(H\to Z'Z')$ is proportional to $m_H^3$ while $\Gamma(H\to NN)$ is proportional to $m_H$.

\section{Heavy Higgs-Strahlung at Muon Collider}\label{SEC:CS}

The novel lepton number violation signatures are induced through the heavy Higgs-strahlung process at the muon collider
\begin{equation}
	\mu^+\mu^-\to Z' H.
\end{equation}

The cross section of this heavy Higgs-strahlung is calculated as
\begin{equation}\label{Eqn:ZpH}
	\sigma(\mu^+\mu^-\to Z' H) = \frac{g'^4}{48\pi s^2} \frac{\lambda(s,m_{Z'}^2,m_H^2)+12 s m_{Z'}^2}{(s-m_{Z'}^2)^2+m_{Z'}^2\Gamma_{Z'}^2}\sqrt{\lambda(s,m_{Z'}^2,m_H^2)},
\end{equation}
where $\Gamma_{Z'}$ is the total decay width of $Z'$, and the function $\lambda(x,y,z)$ is defined as
\begin{equation}
	\lambda(x,y,z)=x^2+y^2+z^2-2xy-2xz-2yz.
\end{equation}

\begin{figure}
	\begin{center}
		\includegraphics[width=0.45\linewidth]{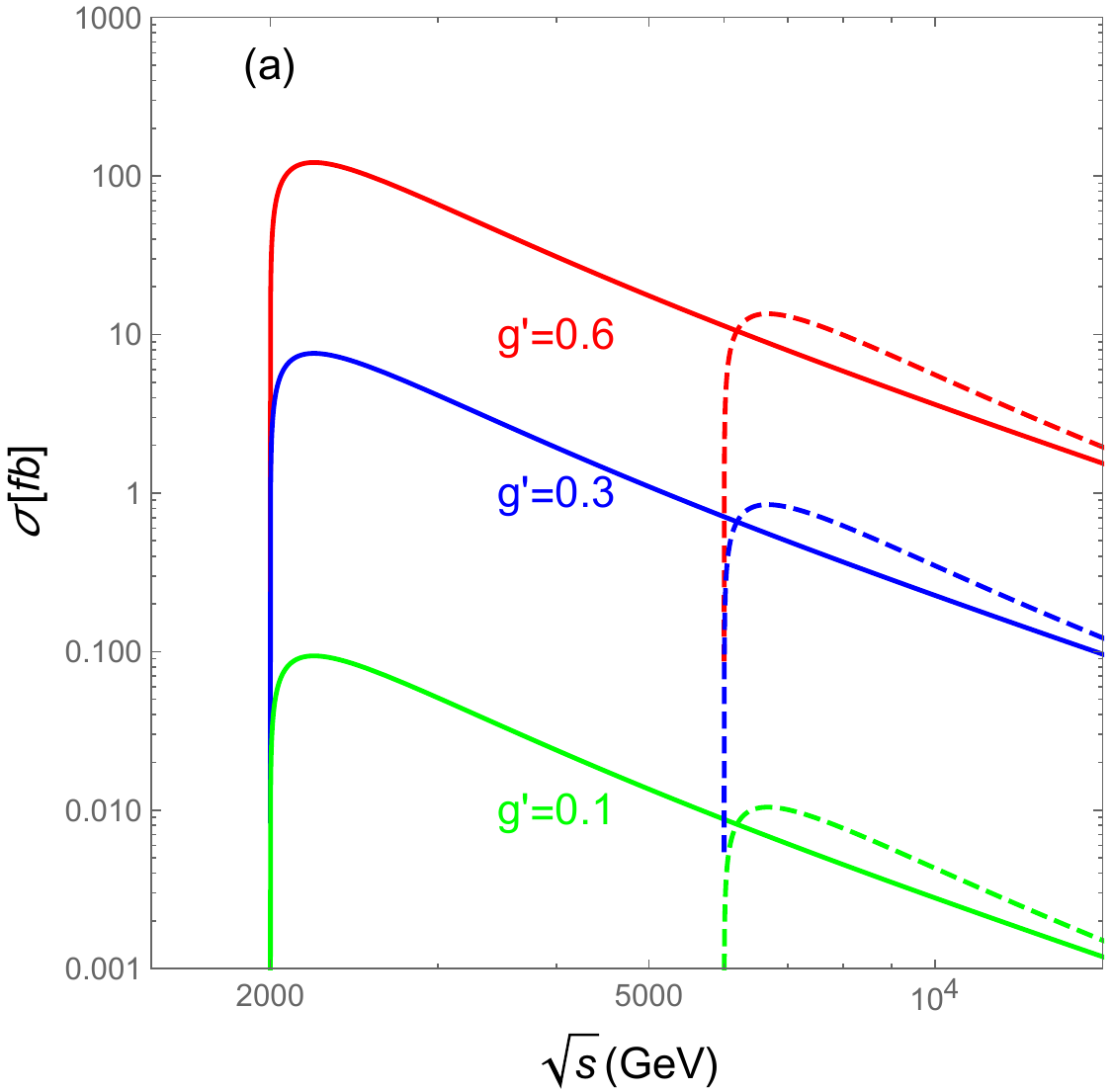}
		\includegraphics[width=0.45\linewidth]{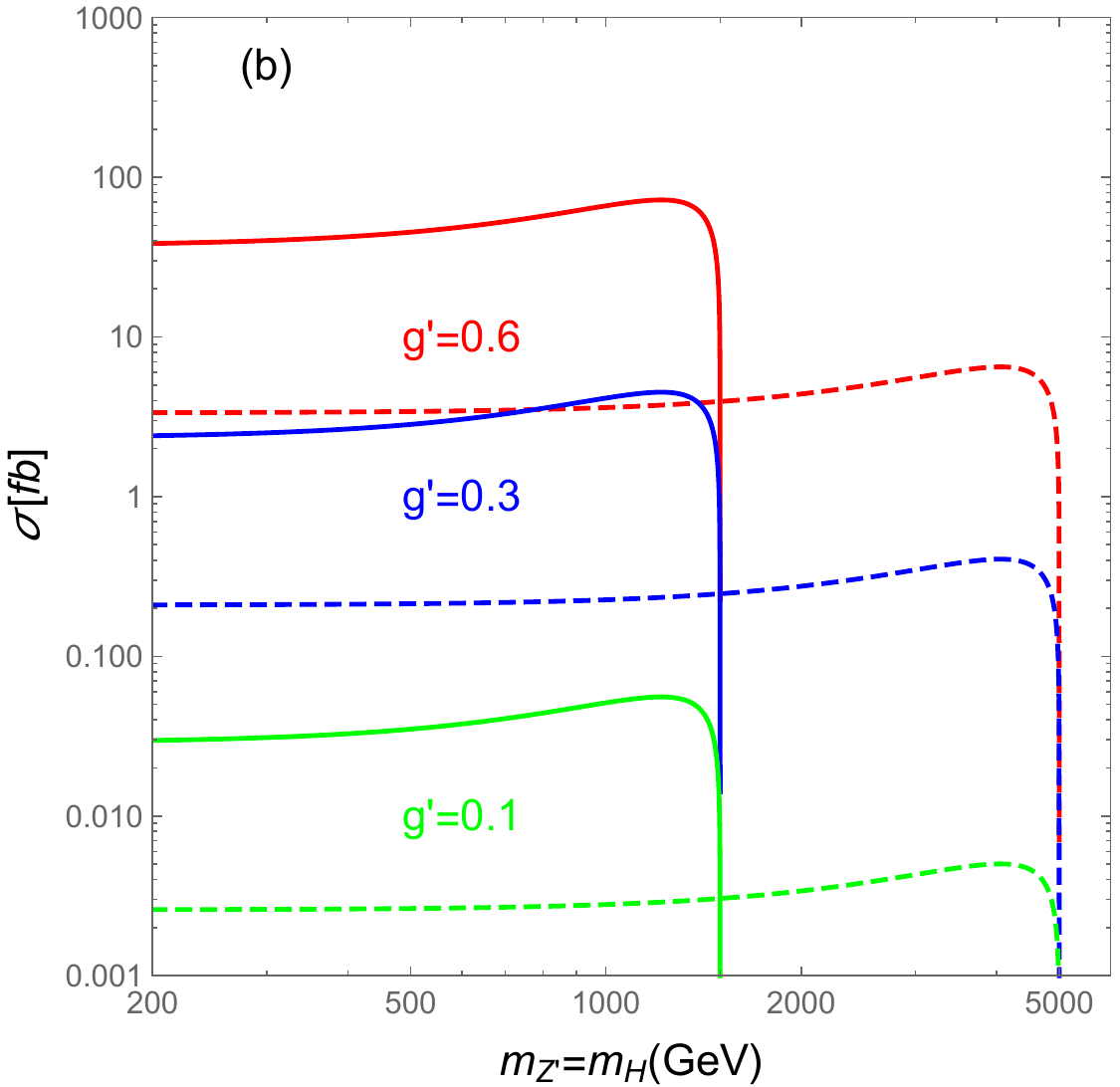}
		\includegraphics[width=0.45\linewidth]{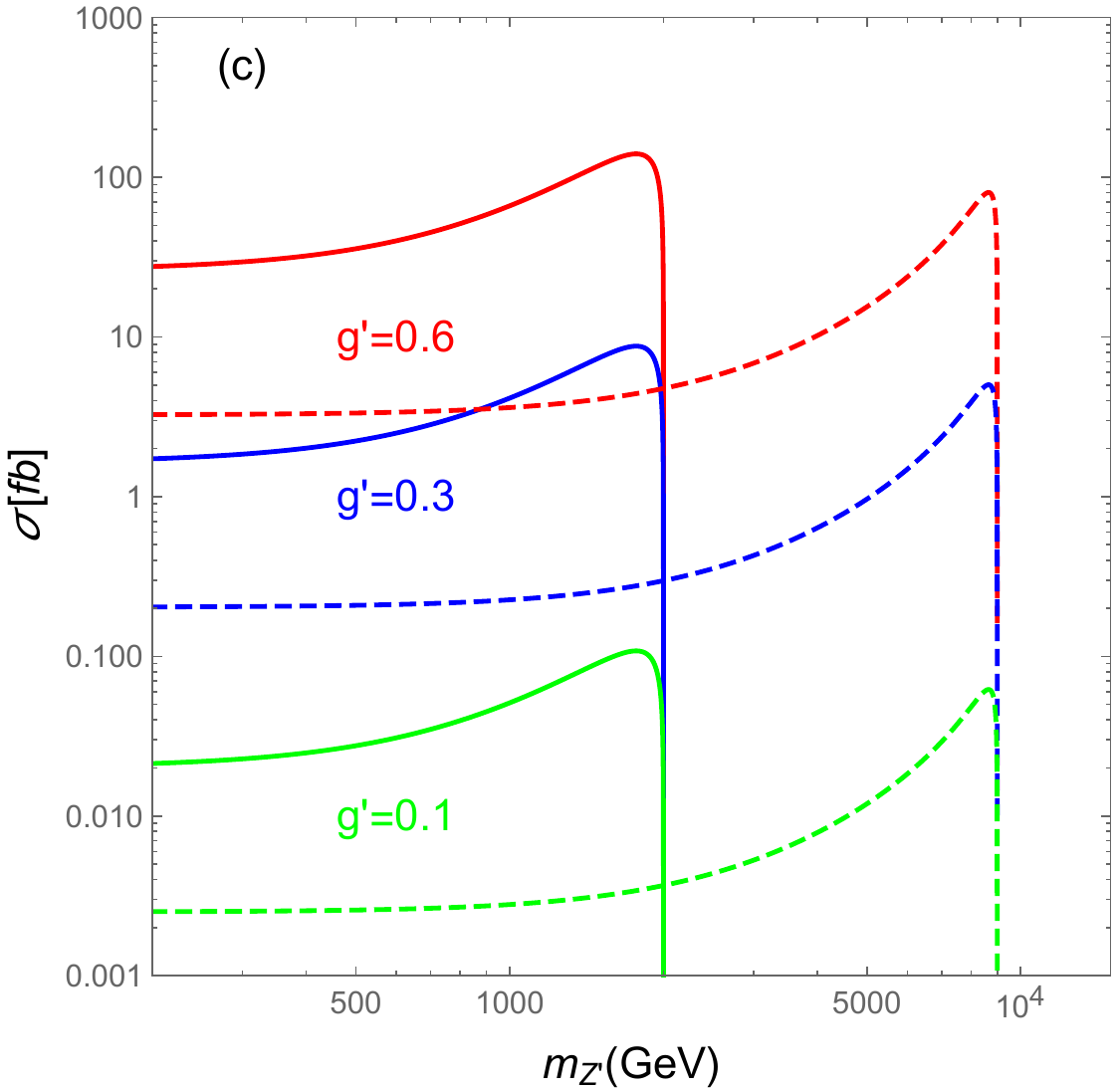}
		\includegraphics[width=0.45\linewidth]{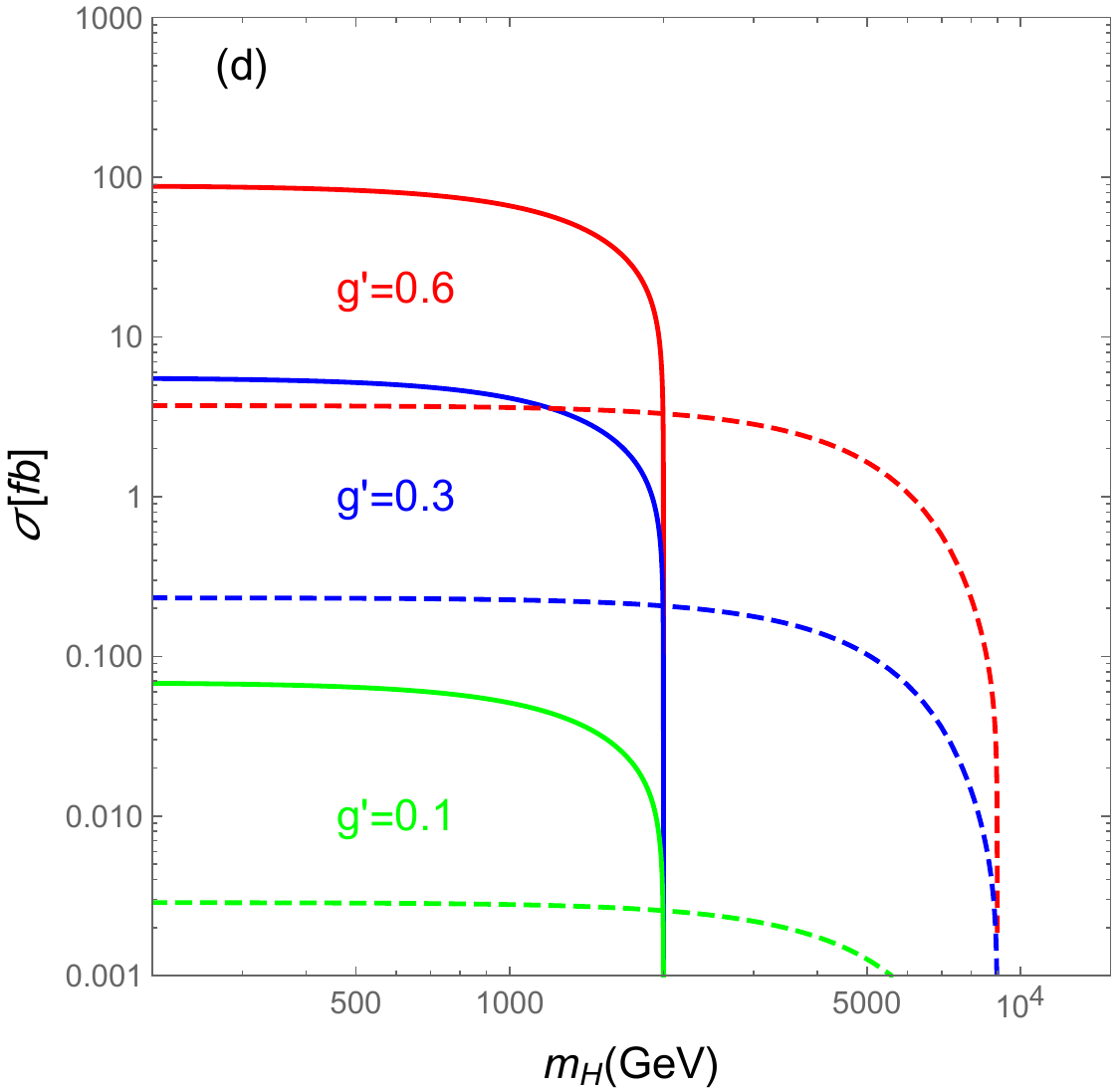}
	\end{center}
	\caption{Cross section of the heavy Higgs-strahlung at muon collider. The red, blue, and green lines are the results for $g'=0.6,0.3,0.1$, respectively. In panel (a), the solid (dashed) lines are the results for $m_{Z'}=m_H=1000 ~(3000)$~GeV. In panel (b), we assume $m_{Z'}=m_H$. In panel (c), we fix $m_H=1000$ GeV. In panel (d), we fix $m_{Z'}=1000$ GeV. Solid (dashed) lines in panel (b)-(d) are the results at the 3 (10) TeV muon collider.}
	\label{fig2}
\end{figure}

In Figure \ref{fig2}, we depict the cross section of heavy Higgs-strahlung for some benchmark scenarios. In panel (a) of Figure \ref{fig2}, we report the cross section as a function of the collision energy for the scenario with $m_{Z'}=m_H=1000(3000)$ GeV. It is clear that the cross section quickly reaches the maximum value when the collision energy is slightly above the threshold $\sqrt{s}>m_{Z'}+m_H$, and decreases as $\sim g'^4/s$ at high energies. 

In panels (b) to (d) of Figure \ref{fig2}, we fix the collision energy of the muon collider to 3 TeV and 10 TeV \cite{Accettura:2023ked,InternationalMuonCollider:2025sys} for illustration. The scenario with $m_{Z'}=m_H$ is shown in panel (b) of Figure \ref{fig2}. We find that apart from the threshold region, the cross section of $\mu^+\mu^-\to Z'H$ increases as $m_{Z'}(=m_H)$ becomes larger. Maximum cross section is obtained with $m_{Z'}=m_H\sim 0.4\sqrt{s}$ in this scenario. Typically, for $g'\gtrsim0.1$, the cross section of $\mu^+\mu^-\to Z'H$ is larger than $10^{-3}$ fb, which is promising with an integrated luminosity larger than 1 ab$^{-1}$ at the muon collider. Meanwhile, with a relatively larger cross section for a fixed value of $g'$, the 3 TeV muon collider is actually more promising than the 10 TeV muon collider to test $Z'$ and $H$ below the TeV scale.

In panel (c) of Figure \ref{fig2}, we fix $m_H=1000$ GeV to show the impact of $m_{Z'}$. Compared to the scenario $m_{Z'}=m_H$, increasing $m_{Z'}$ alone would greatly enhance the cross section near the threshold, which is due to the cancellation in the denominator of Equation \eqref{Eqn:ZpH}. On the other hand, increasing $m_H$ alone would lead to a smaller cross section, which is shown in panel (d) of Figure \ref{fig2}.

\allowdisplaybreaks

With vanishing Higgs mixing angle $\sin\alpha=0$, the decay $H\to NN$ is the only viable channel of heavy Higgs with $m_{Z'}=m_H$. When the new gauge boson also decays into heavy neutral leptons $Z'\to NN$, there are a total of four heavy neutral leptons in the final states. Further decay of the heavy neutral leptons could induce various interesting signatures. Considering the fully visible decay mode $N\to \ell^\pm W^\mp \to \ell^\pm J$, the promising signatures from $4N$ are
\begin{eqnarray}
	4N &\to& 4\ell^\pm +4J, \\
	  &\to & 3\ell^\pm + \ell^\mp + 4J, \\
	  &\to & 2\ell^+ + 2\ell^- +4J.
\end{eqnarray}

On the other hand, when considering the leptonic decay of the new gauge boson $Z'\to \ell^+\ell^-,\nu_\ell \nu_\ell$, the interesting signatures would be 
\begin{eqnarray}
	\ell^+\ell^- NN &\to& \ell^+ \ell^- + \ell^\pm J + \ell^\pm J \to 3\ell^\pm + \ell^\mp +2J \\
	  &\to& \ell^+ \ell^- + \ell^+ J + \ell^- J \to 2\ell^+ + 2 \ell^- +2J \\
	\nu_\ell \nu_\ell NN  &\to&\nu_\ell \nu_\ell + \ell^\pm J + \ell^\pm J \to 2\ell^\pm +2J +\cancel{E}_T \\
	&\to& \nu_\ell \nu_\ell + \ell^+ J + \ell^- J \to \ell^+  \ell^- +2J+ \cancel{E}_T
\end{eqnarray}

An investigation of all these signatures from the cascade $Z'H$ decay is beyond the scope of this paper. In the following, we study the novel same-sign tetralepton signature $4\ell^\pm +4J$ and the same-sign trilepton signature $3\ell^\pm +\ell^\mp +2J$. These two lepton number violation signatures have a relatively clean background process. Meanwhile, the heavy resonances $Z'$, $H$, and $N$ can all be reconstructed as we consider the fully visible decay chain.

\section{Same-Sign Tetralepton Signature}\label{SEC:SG1}

In this paper, we consider a muon-flavored heavy neutral lepton for illustration, i.e., $V_{\mu N}\neq0, V_{e N}=V_{\tau N}=0$. For simplicity, we assume $m_{Z'}=m_H$ in the following studies. The full process of the same-sign tetralepton signature is 
\begin{equation}
	\mu^+\mu^-\to Z' H \to NN + NN \to 4 \mu^\pm +4 W^\mp \to 4\mu^\pm +4J,
\end{equation}
where the fat-jets $J$ come from the hadronic decay of $W$ bosons. There are no irreducible backgrounds of this novel signature. In the following studies, we assume one background event $N_B=1$ for a conservative estimation.

\begin{figure}
	\begin{center}
		\includegraphics[width=0.45\linewidth,height=0.3\linewidth]{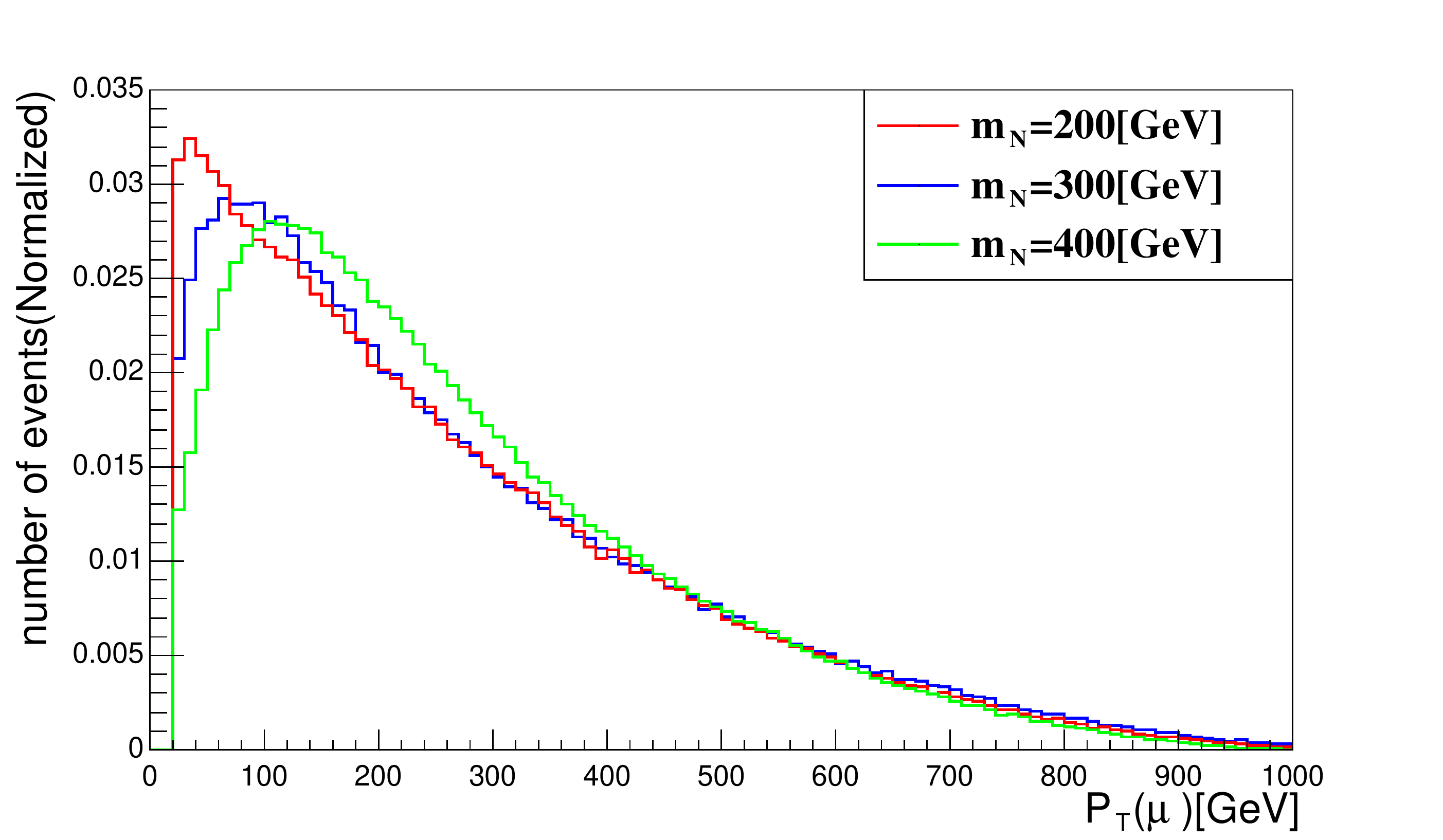}
		\includegraphics[width=0.45\linewidth,height=0.3\linewidth]{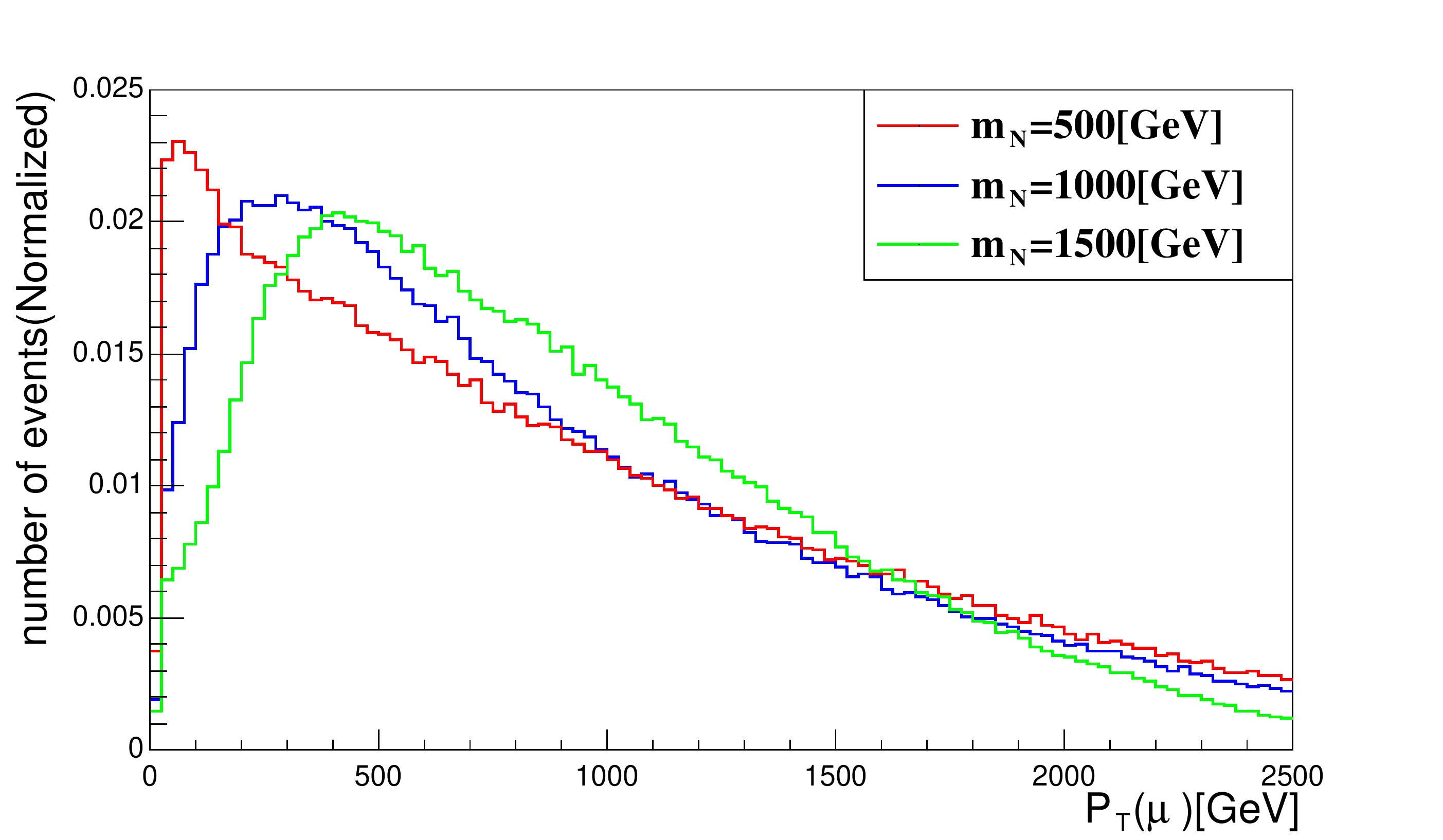}
		\includegraphics[width=0.45\linewidth,height=0.3\linewidth]{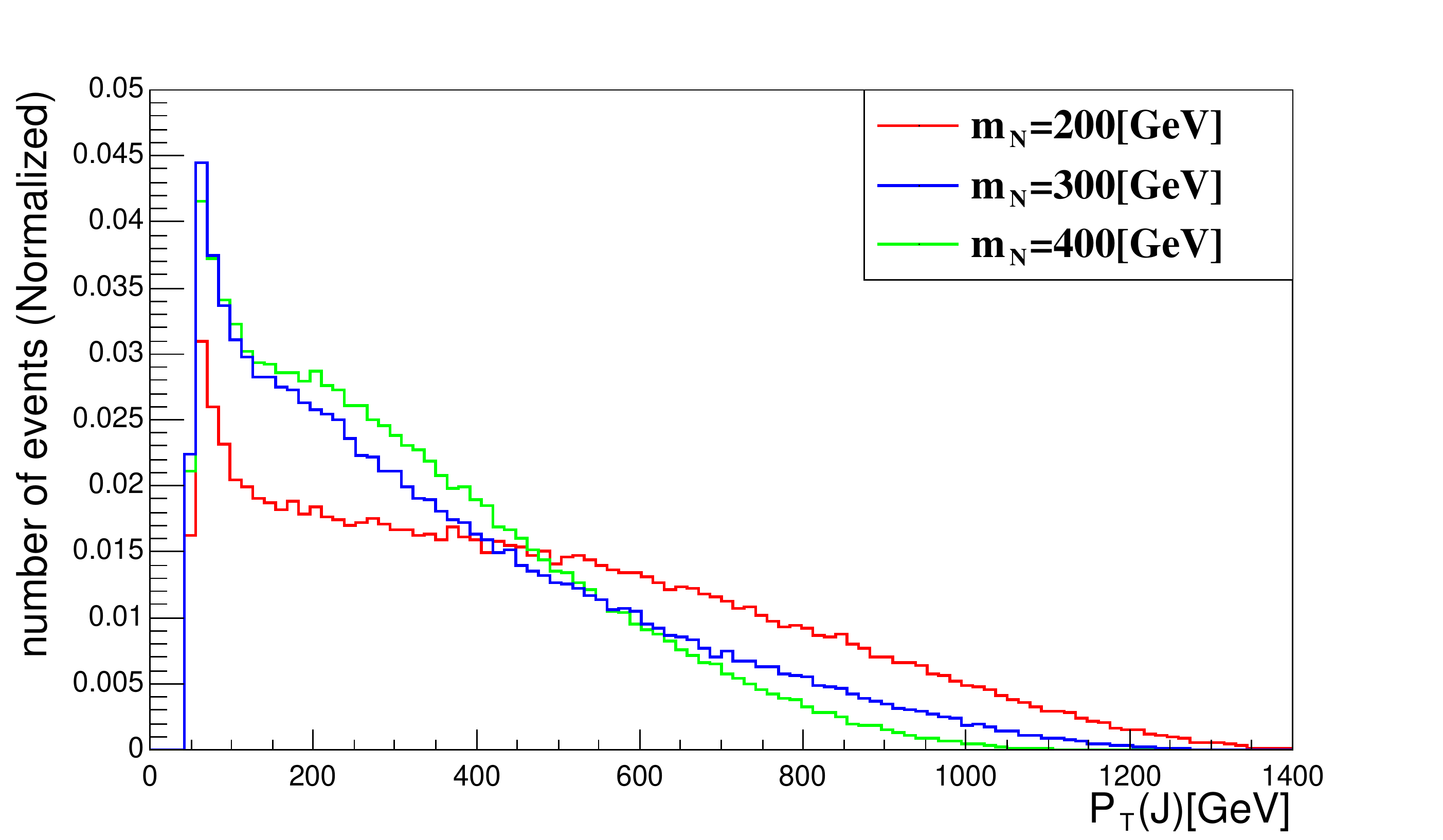}
		\includegraphics[width=0.45\linewidth,height=0.3\linewidth]{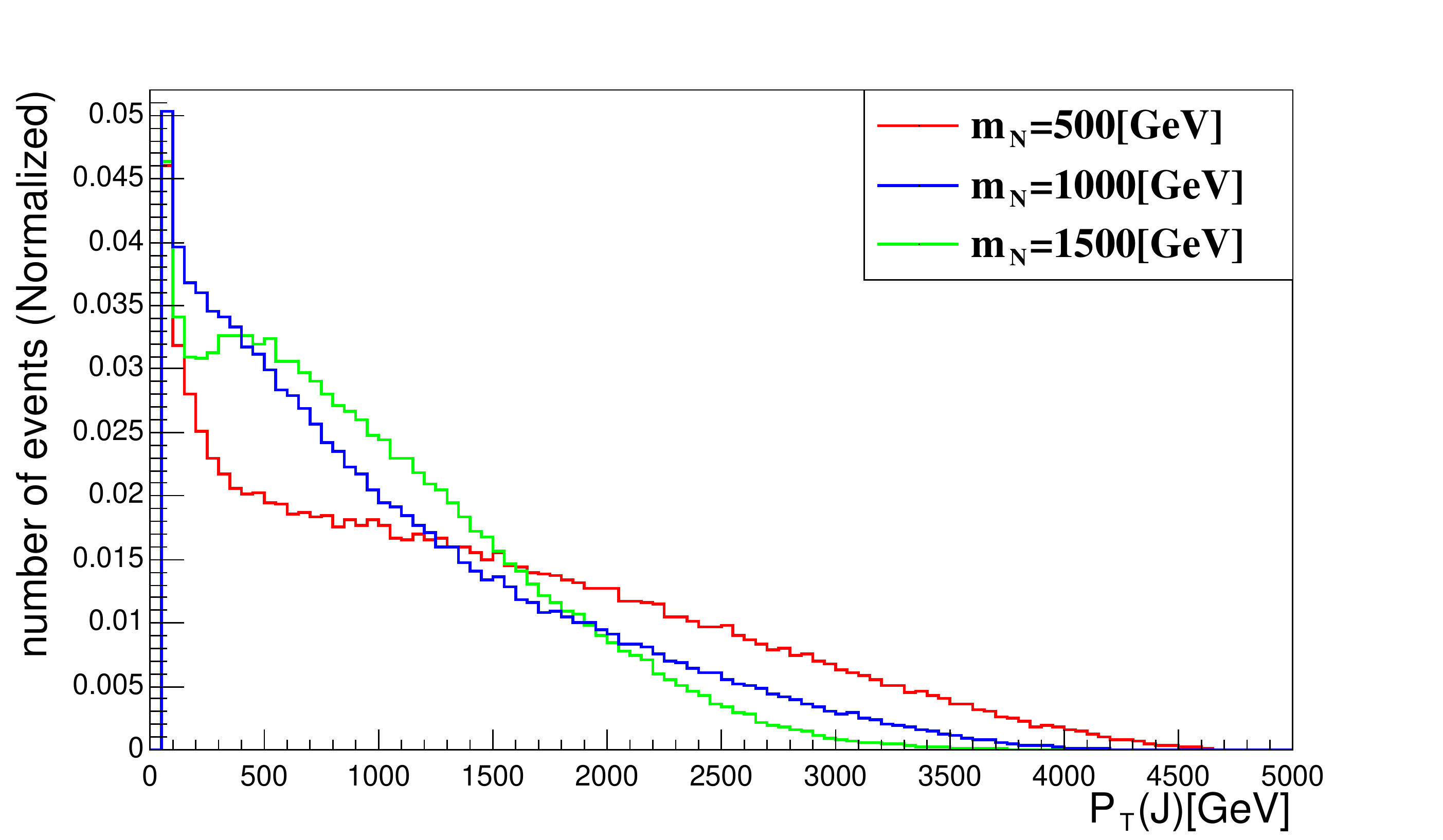}
		\includegraphics[width=0.45\linewidth,height=0.3\linewidth]{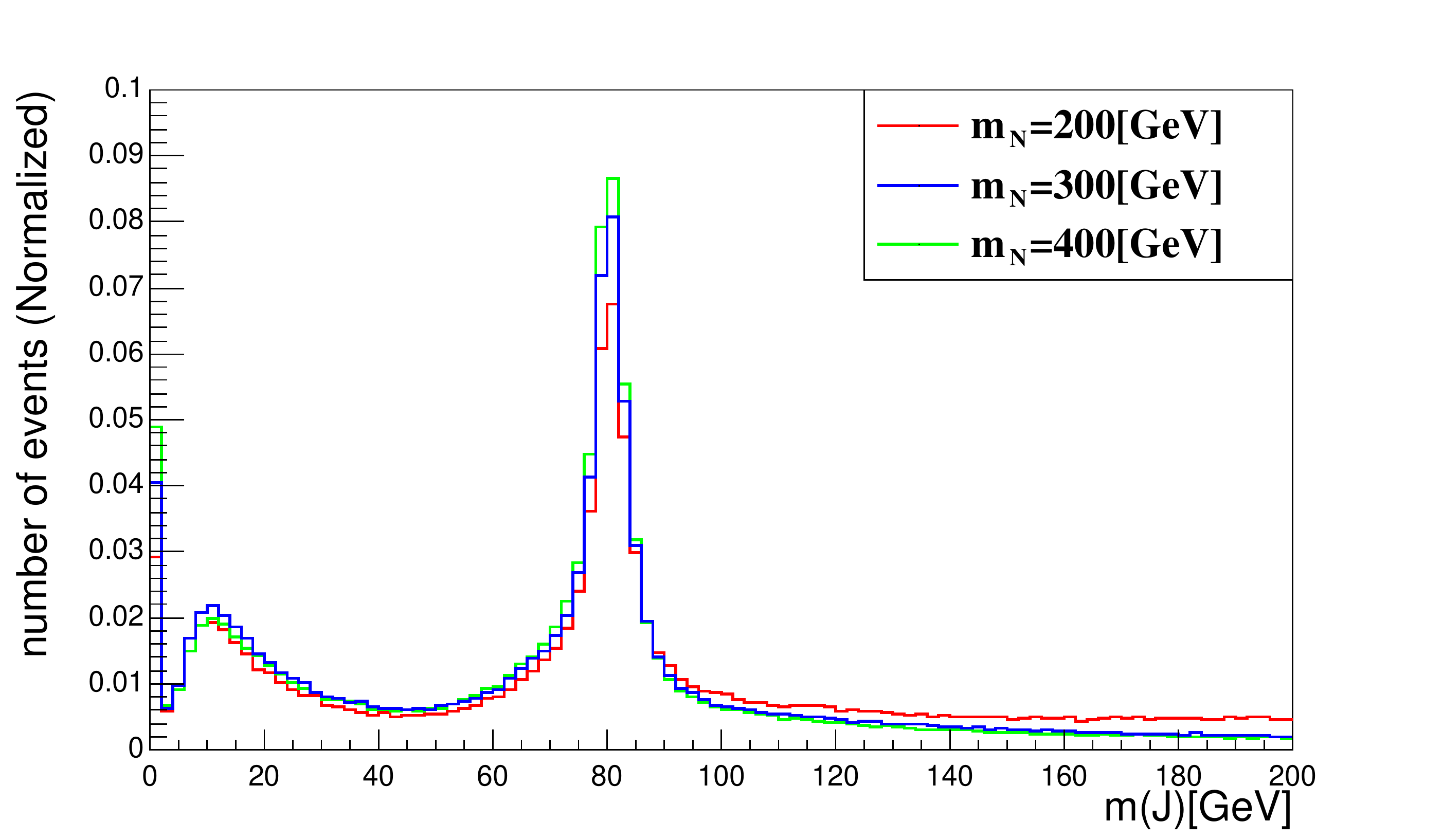}
		\includegraphics[width=0.45\linewidth,height=0.3\linewidth]{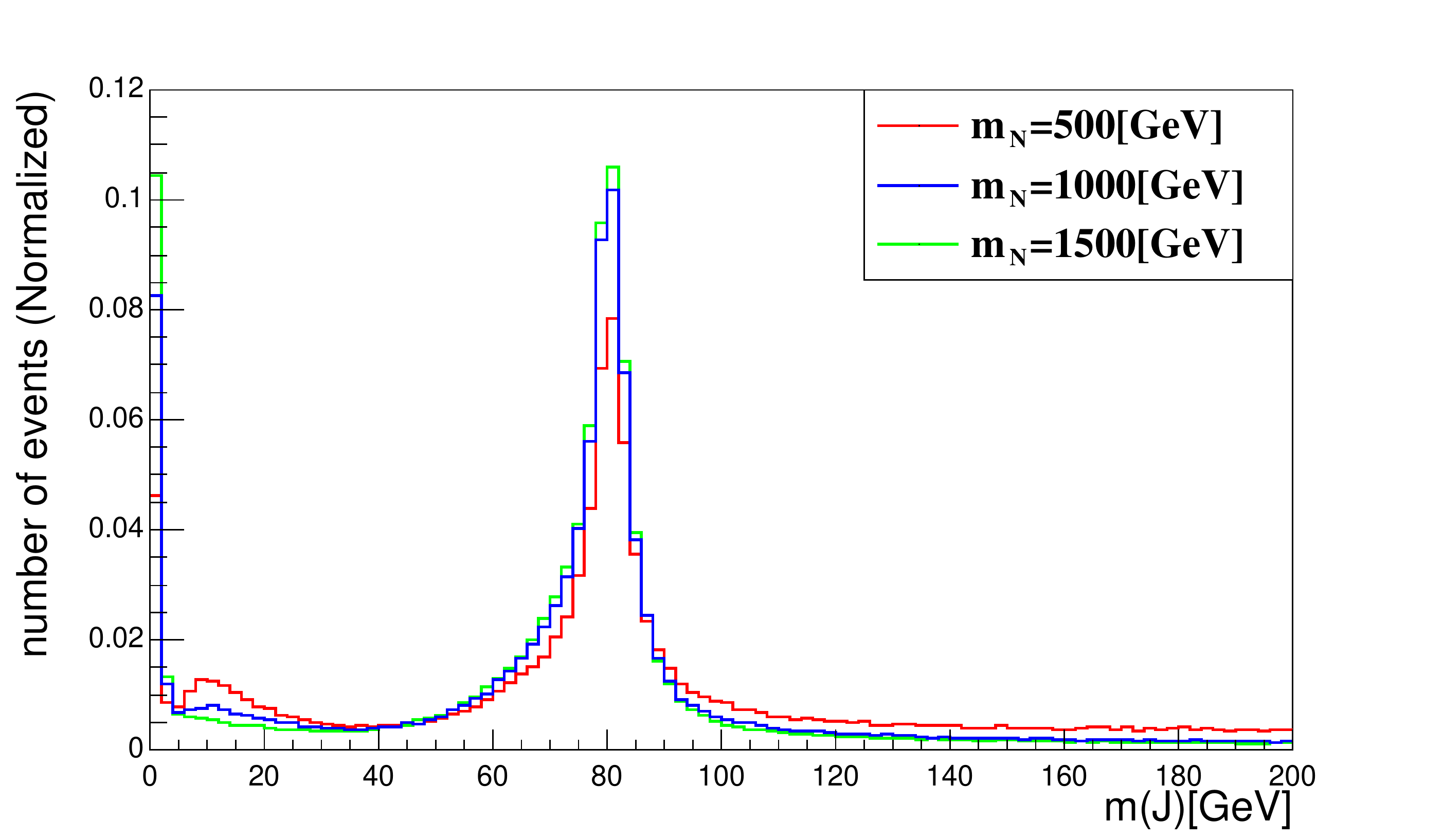}
	\end{center}
	\caption{Normalized distributions of muon transverse momentum $P_T(\mu)$, fat-jet transverse momentum $P_T(J)$, and invariant mass of fat-jet $m_J$. Left panels are the results of 3 TeV, and right panels are the results of 10 TeV. During the simulation of these plots, we have assumed the relation $m_{Z'}=m_H=3m_N$ for illustration.
		\label{fig3}}
\end{figure}

In the simulation, we use the {\bf FeynRules}  package \cite{Alloul:2013bka} to implement the $U(1)_{L_\mu-L_\tau}$ extension of the type-I seesaw model. We then use {\bf Madgraph5\_aMC@NLO} \cite{Alwall:2014hca} to generate the  leading order  signal events at the parton level. After this, {\bf Pythia8} \cite{Sjostrand:2014zea} is used to do parton showering and hadronization, and the detector simulation is performed by {\bf Delphes3} \cite{deFavereau:2013fsa} with the corresponding muon collider card. The fat-jets are reconstructed by using the Valencia algorithm \cite{Boronat:2014hva} with $R = 1.2$.

Since this signature is background free, we apply the default pre-selection cuts of {\bf Delphes3}:
\begin{align}\label{Eqn:ps}
P_T(\mu)>20~{\rm GeV}, |\eta(\mu)|<2.5, P_T(J)>50~{\rm{GeV}}, |\eta(J)|<2.5.
\end{align}
  For the muon isolation criterion, we use the built-in muon collider strategy: inside a cone of $\Delta R=0.1$, the ratio of the scalar sum of $P_T$ of all other particles to $P_T(\mu)$  is smaller than 0.2. 
The distributions of some variables for the signal at the 3 TeV and 10 TeV muon collider after the pre-selection cuts are shown in Figure \ref{fig3}. It is obvious that with a larger heavy neutral lepton mass $m_N$, the transverse momentum of the final state muon $P_T(\mu)$ tends to be more energetic. We also report that the Valencia fat-jet algorithm is more efficient for reconstructing fat-jet from heavier $N$.

In order to reconstruct the heavy resonance through the decay chain $H/Z'\to NN\to \mu^\pm \mu^\pm JJ$, it is enough to require two fat-jets.  For the gauge boson, the purely leptonic channel $Z'\to \mu^+\mu^-$ is more promising due to higher detection efficiency, which will be considered in Section \ref{SEC:SG2}.  The same-sign tetralepton signature is then selected as
\begin{align}
	N(\mu^\pm)=4, N(J)\geq 2.
\end{align} 
where at least two fat-jets are required in order to keep more signal events during the analysis. According to Figure \ref{fig3}, we also require the fat-jet mass within the range of 
\begin{equation}
	50~\text{GeV}<m_J<110~\text{GeV},
\end{equation}
to make sure it's coming from the decay of the $W$ boson.  Since this signature is background free, no further cut is applied.

\begin{table}
	\begin{center}
		\begin{tabular}{c | c | c | c | c |c} 
			\hline
			\hline
			\multicolumn{2}{c|}{$\mu^+\mu^-\to 4\mu^\pm+4J$} & Pre-Selection (fb) & After Selection (fb) & Significance & $5\sigma$ Luminosity (fb$^{-1}$)\\
			\hline
			\multirow{3}{*}{3 TeV}	& $m_N=200$ GeV & $7.33\times10^{-2}$ & $3.09\times10^{-2}$ & 12.6 & 269.3 \\			
			\cline{2-6}
			& $m_N=300$ GeV & $8.80\times10^{-2}$ & $4.84\times10^{-2}$ & 17.0 & 171.9 \\			
			\cline{2-6}
			& $m_N=400$ GeV & $1.01\times10^{-1}$ & $5.56\times10^{-2}$ & 18.6 & 149.7 \\			
			\hline
			\multirow{3}{*}{10 TeV}	& $m_N=500$ GeV & $5.56\times10^{-3}$  & $2.35\times10^{-3}$ & 10.5 & 3541 \\			
			\cline{2-6}
			& $m_N=1000$ GeV & $7.82\times10^{-3}$ & $4.21\times10^{-3}$ & 15.5 & 1978 \\			
			\cline{2-6}
			& $m_N=1500$ GeV & $8.07\times10^{-3}$ & $4.33\times10^{-3}$ & 15.8 & 1925 \\			
			\hline
			\hline
		\end{tabular}
	\end{center}
	\caption{ Results for the benchmark points at the muon collider, where we have fixed the relation $m_{Z'}=m_H=3 m_N$ and $g'=0.6$. The significance is calculated with an integrated luminosity of 1 (10) ab$^{-1}$ for the 3 TeV (10 TeV) muon collider. }
	\label{Tab01}
\end{table}

In Table \ref{Tab01}, we show the results for some benchmark points at the muon collider.
The significance is calculated as~\cite{Cowan:2010js}
\begin{align}
	S=\sqrt{2\left[(N_S+N_B)\ln\left(1+\frac{N_S}{N_B}\right)-N_S\right]},
\end{align}
where $N_S$ and $N_B$ are the events number of the signal and background, respectively. The benchmark points for the 3 TeV muon collider are selected as $m_N=$200, 300, and 400 GeV with the mass relation $m_{Z'}=m_H=3 m_N$ and $g'=0.6$. After all selection cuts, the cross sections are $3.09\times10^{-2}$ fb, $4.84\times10^{-2}$ fb, and $5.56\times10^{-2}$~fb for $m_N=$200 GeV, 300 GeV, and 400 GeV, respectively. The cross sections clearly increase as $m_N$ becomes larger, which means that the muon collider is more sensitive to heavier $N$. With an integrated luminosity of 1 ab$^{-1}$ at the 3 TeV muon collider, the significance could reach 12.6, 17.0, and 18.6 for $m_N=$200 GeV, 300 GeV, and 400 GeV. The benchmark point with $m_N=200$ GeV can be discovered with the luminosity of 269.3 fb$^{-1}$. Meanwhile, it is enough with 200 fb$^{-1}$ to discover those benchmarks with $m_N=$300 GeV and 400 GeV.

As shown in Figure \ref{fig2}, the cross section of the heavy Higgs-strahlung $\mu^+\mu^-\to Z'H$ at the 3 TeV muon collider is actually larger than it at the 10 TeV muon collider, so the 3 TeV muon collider is more promising to probe relatively light resonances. We then choose the benchmark points as $m_N=$ 500~GeV, 1000~GeV, and 1500 GeV with $m_{Z'}=m_H=3 m_N$ for the 10 TeV muon collider. The production cross sections of these 10 TeV benchmarks are approximately an order of magnitude smaller than those of the 3 TeV ones. After all selection cuts, the cross sections are $2.35\times10^{-3}$ fb, $4.21\times10^{-3}$ fb, and $4.33\times10^{-3}$~fb for $m_N=$500 GeV, 1000 GeV, and 1500~GeV, which leads to the significance of 10.5, 15.5, and 15.8 with an integrated luminosity of 10 ab$^{-1}$, respectively. Therefore, thanks to much higher integrated luminosity, the TeV-scale new particles are also within the reach of the 10 TeV muon collider through the same-sign tetralepton signature.

\begin{figure}
\begin{center}
	\includegraphics[width=0.45\linewidth,height=0.3\linewidth]{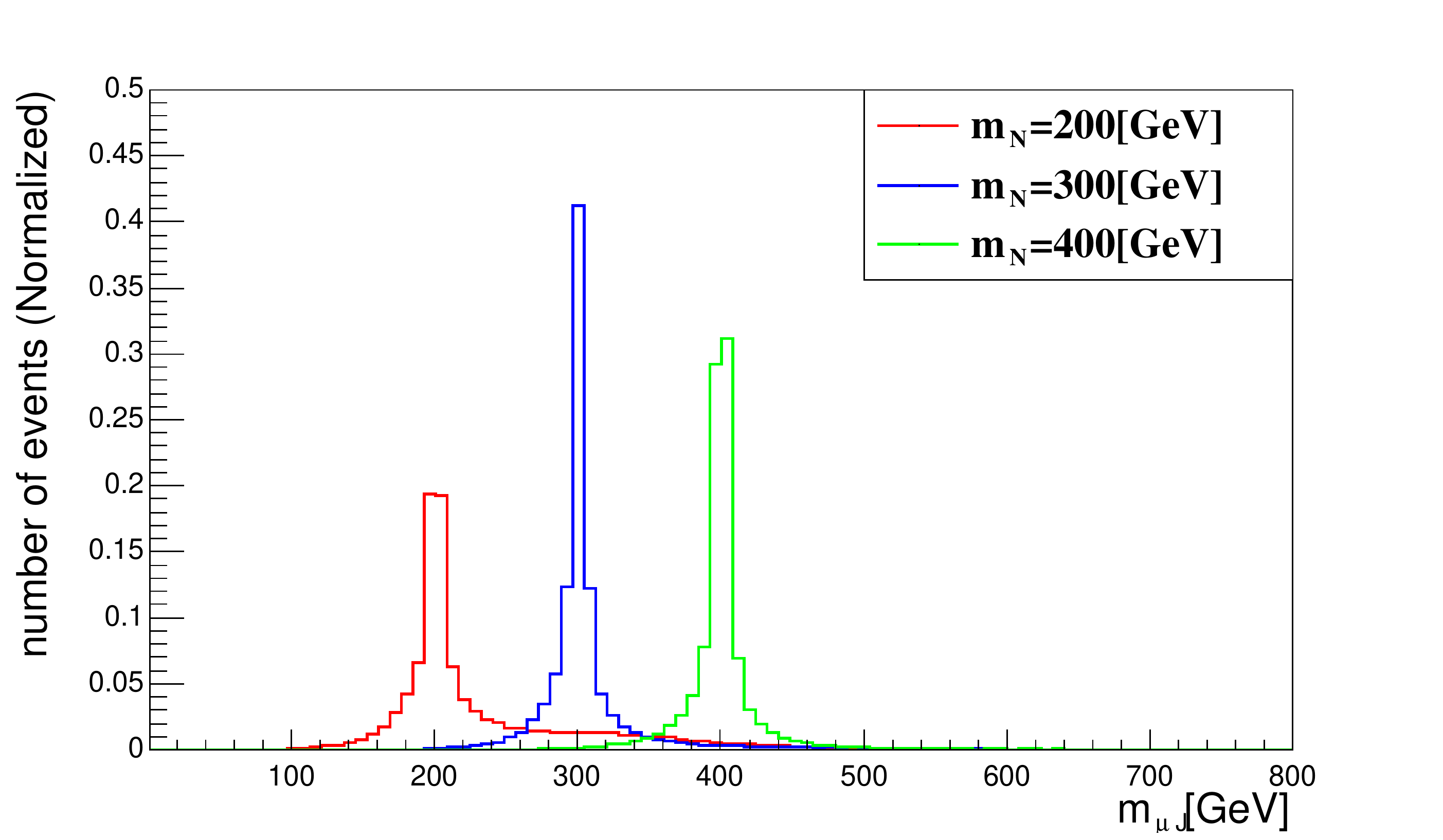}
	\includegraphics[width=0.45\linewidth,height=0.3\linewidth]{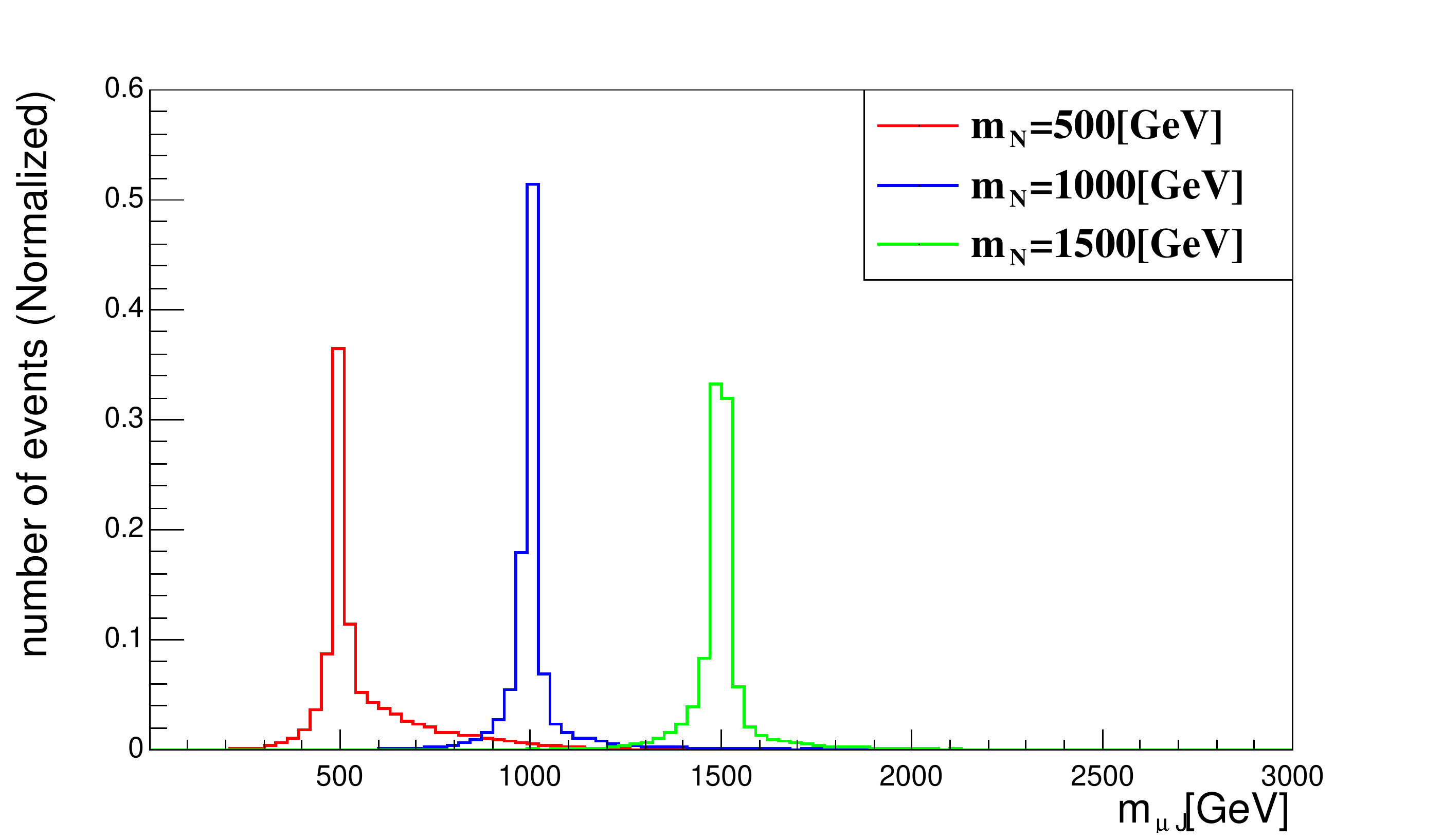}
	\includegraphics[width=0.45\linewidth,height=0.3\linewidth]{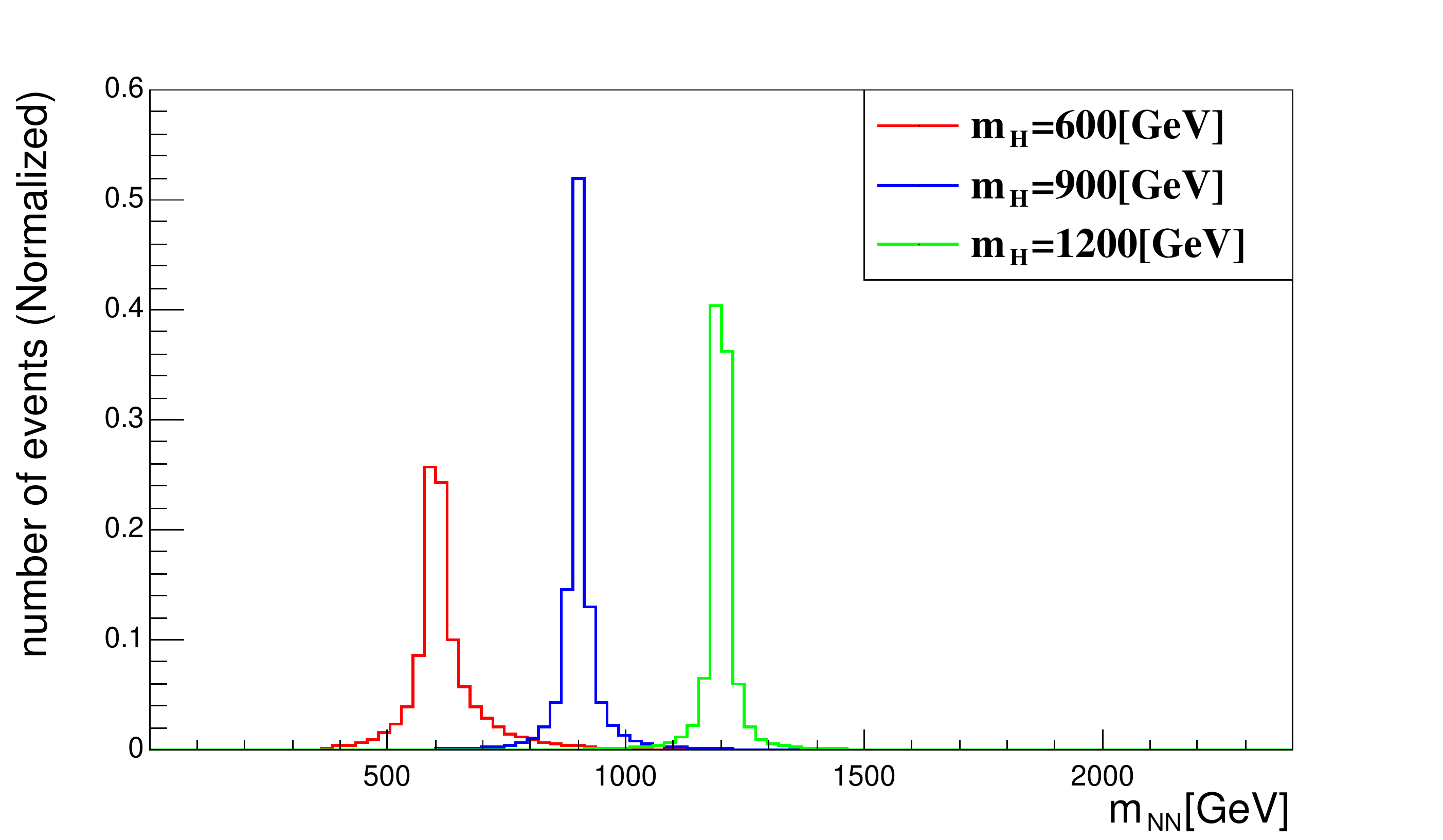}
	\includegraphics[width=0.45\linewidth,height=0.3\linewidth]{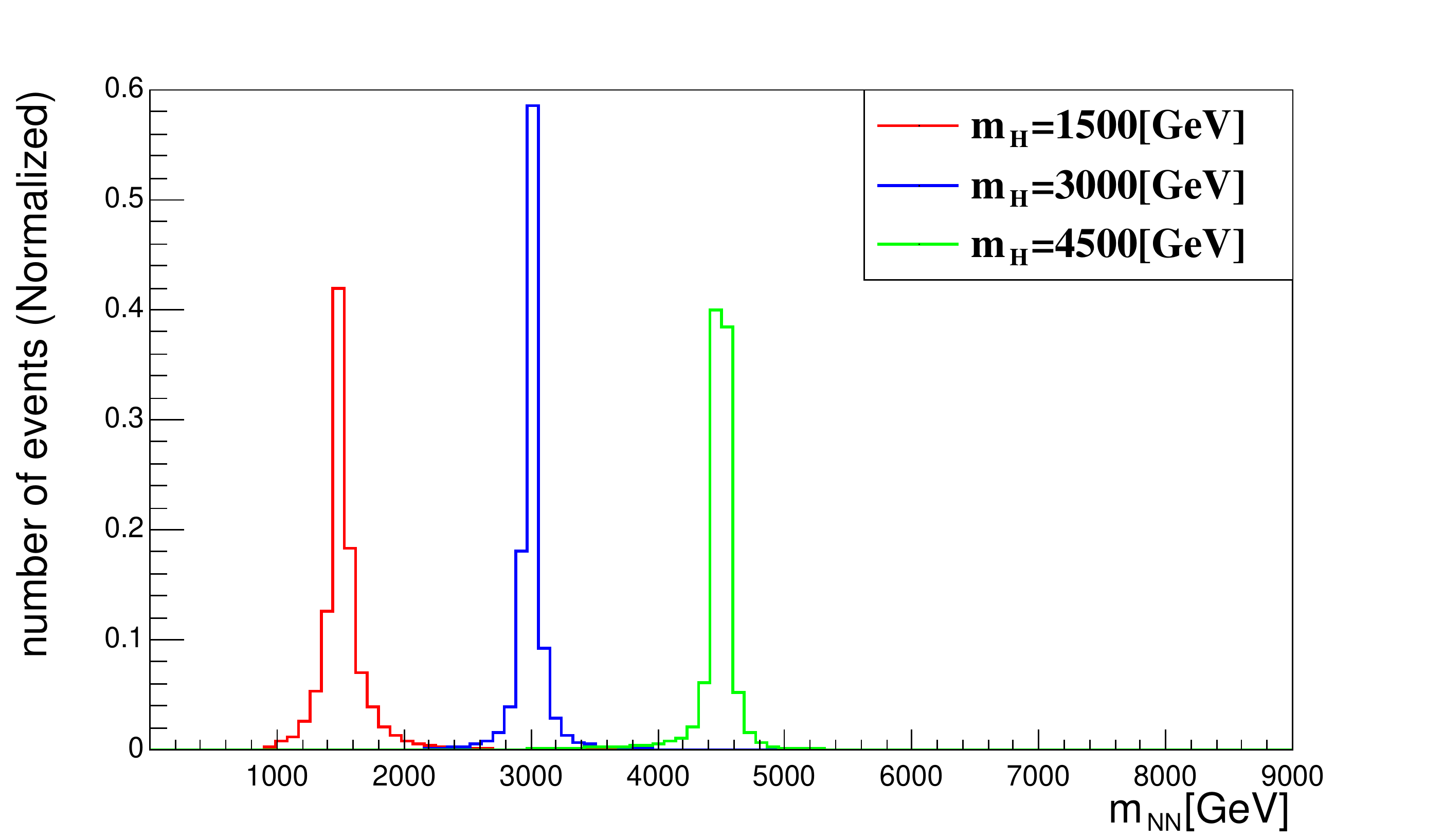}
\end{center}
\caption{Normalized distributions of invariant mass of muon and fat-jet $m_{\mu J}$ and invariant mass of two reconstructed heavy neutral lepton $m_{NN}$. Left panels are the results of 3 TeV, and right panels are the results of 10 TeV. The relation $m_{Z'}=m_H=3m_N$ is also assumed. 
	\label{fig4}}
	\end{figure}

The masses of all new particles can be reconstructed by the final states of the same-sign tetralepton signature. According to the heavy neutral lepton decay $N\to \mu^\pm J$, its mass is determined by the invariant mass of one muon and one fat-jet $m_{\mu J}$. With multi muons and fat-jets in the signature, we reconstruct the heavy neutral lepton masses by minimizing \cite{Liu:2021akf}
\begin{equation}
	\chi^2_N=\sum_{i,j} (m_{\mu^\pm_i J_j}-m_N)^2
\end{equation}
for all detected muons and fat-jets. The reconstructed heavy neutral lepton masses of the benchmark points are shown in Figure \ref{fig4}. To obtain the mass of heavy Higgs or gauge boson through the decay $H/Z'\to NN$, we can consider the invariant mass of the reconstructed heavy neutral leptons.  To determine the correct origin of the heavy neutral leptons, the heavy Higgs or gauge boson is reconstructed by minimizing
\begin{equation} \label{Eqn:chiH}
\chi^2_{HZ'} = \sum_{i,j,k,l} (m_{{N_i}{ N_j}} - m_H)^2+(m_{{N_k}{ N_l}} - m_{Z'})^2 
\end{equation}
for all reconstructed heavy neutral leptons. It should be mentioned that Equation \eqref{Eqn:chiH} is also suitable for the general scenario with $m_{H}\neq m_{Z'}$. The results are shown in Figure \ref{fig4}, where only one resonance appears for a certain benchmark point.

\begin{figure}
	\begin{center}
		\includegraphics[width=0.45\linewidth,height=0.3\linewidth]{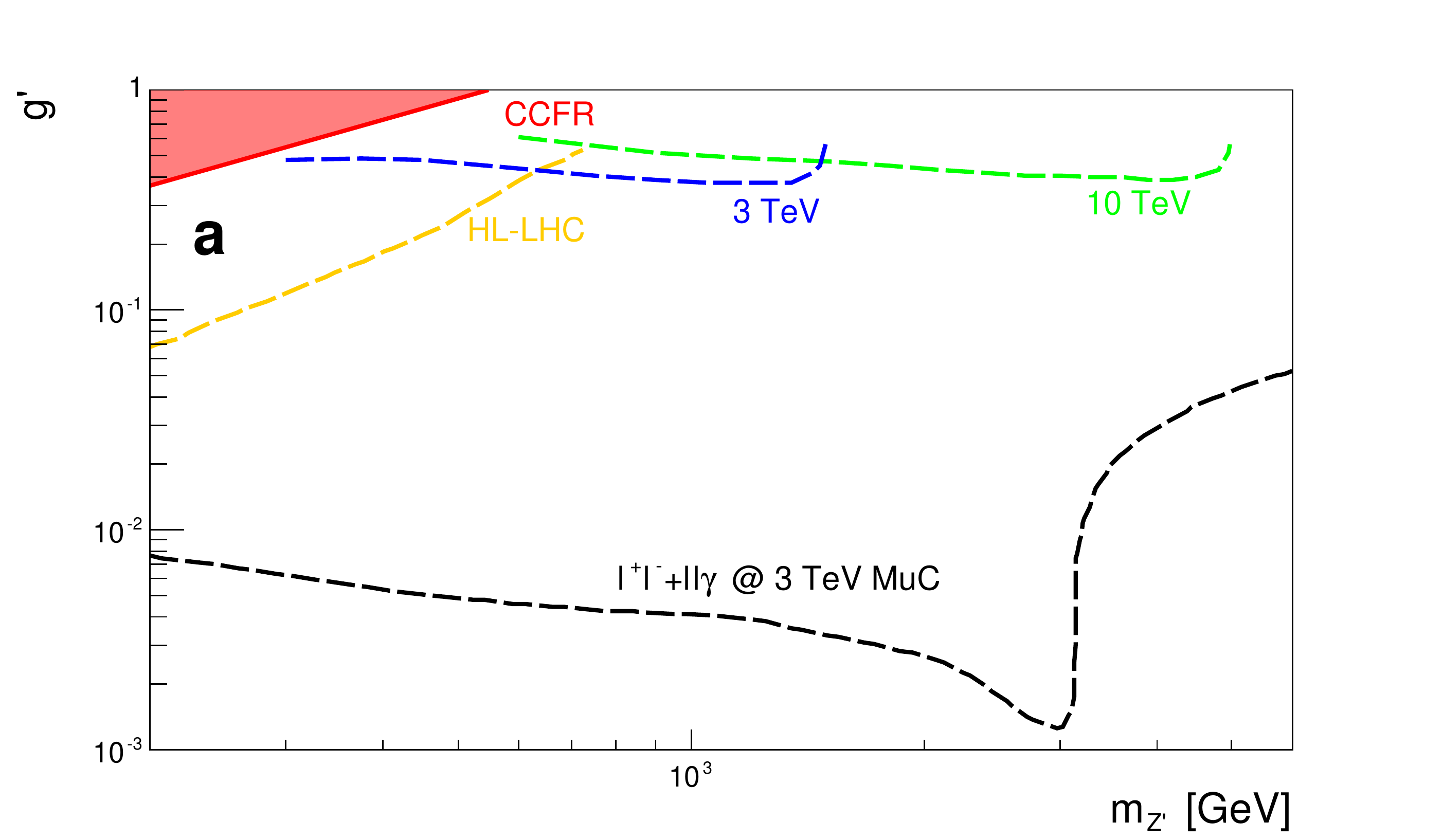}
		\includegraphics[width=0.45\linewidth,height=0.3\linewidth]{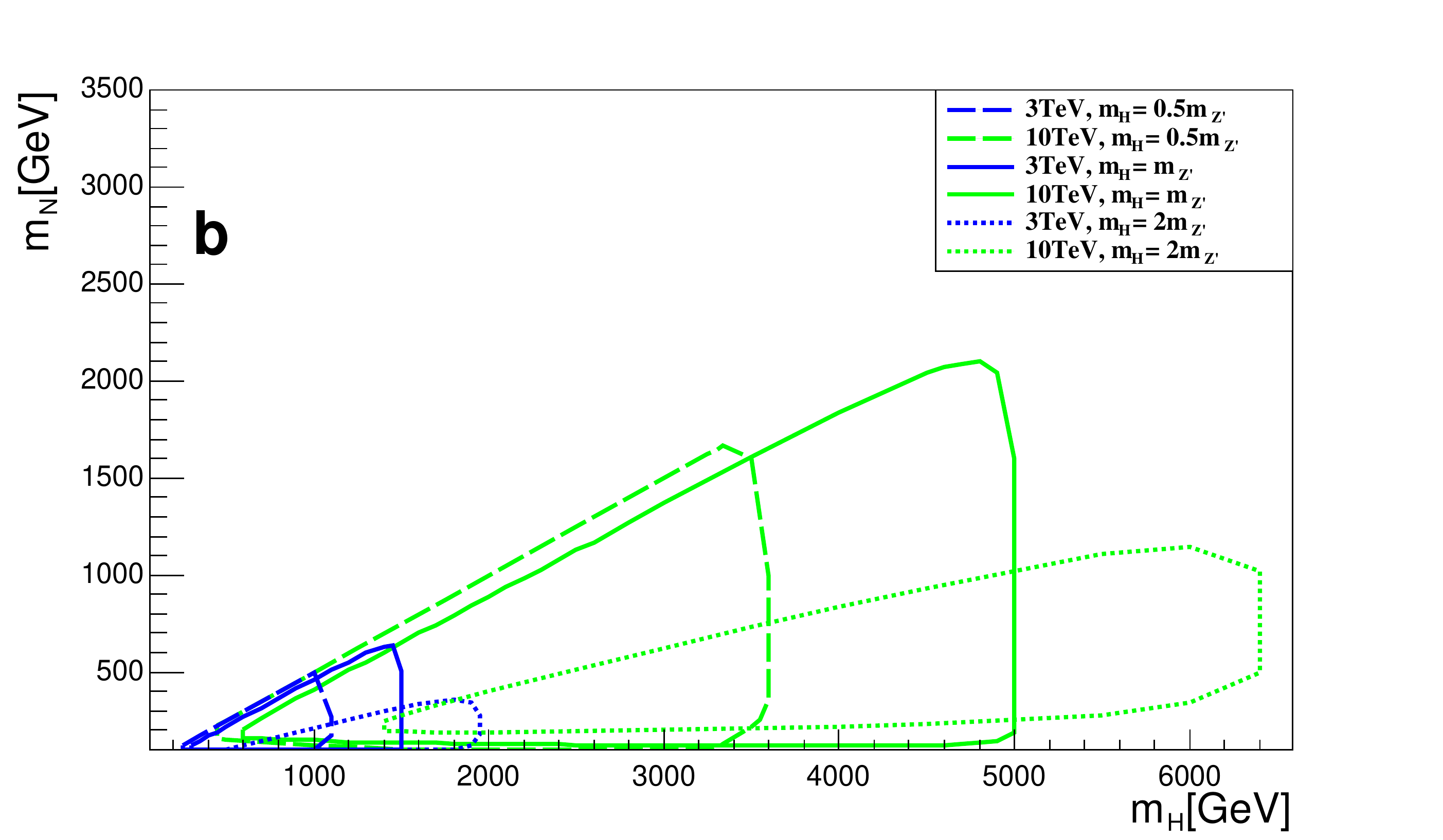}
		\includegraphics[width=0.45\linewidth,height=0.3\linewidth]{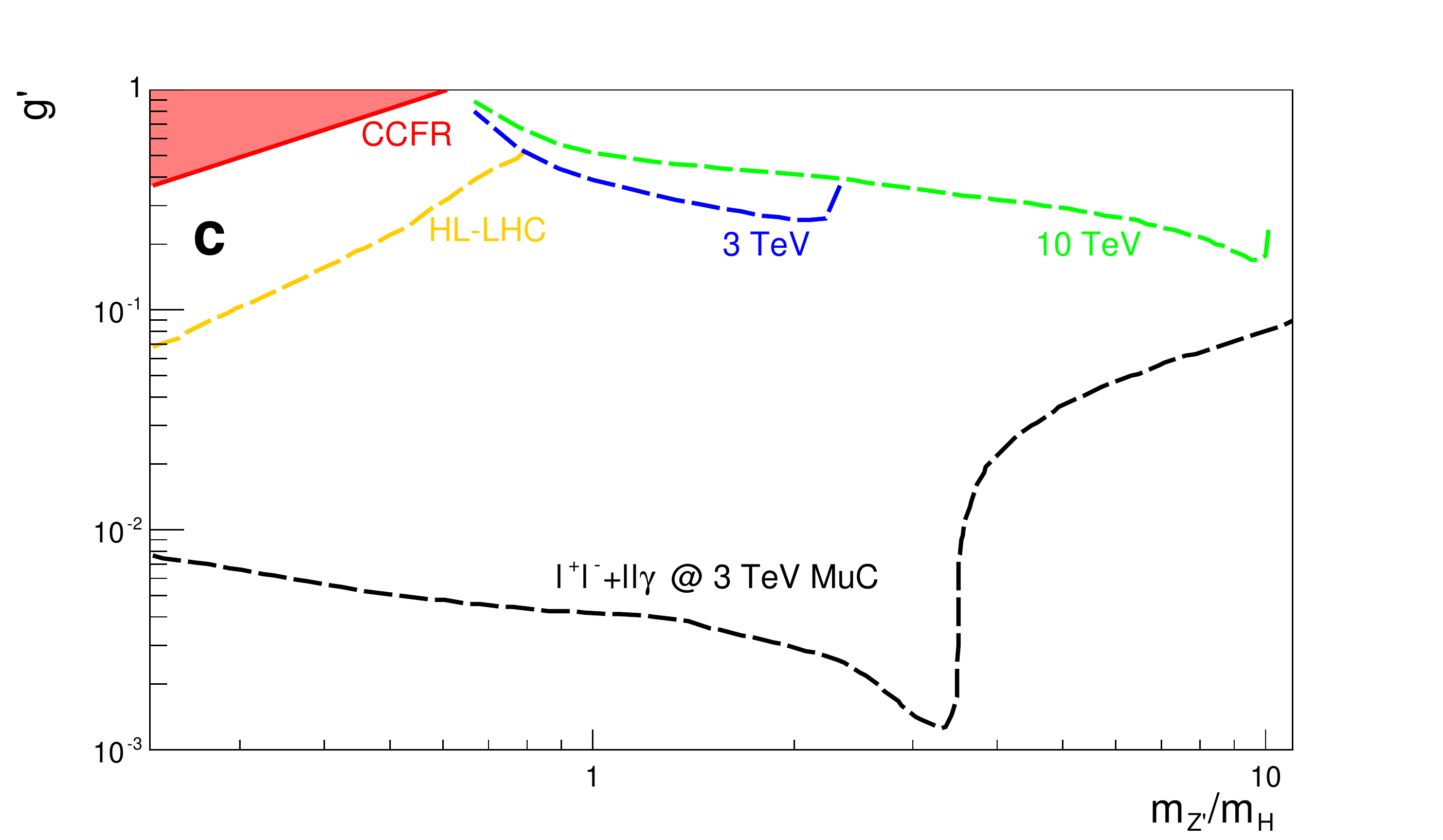}
		\includegraphics[width=0.45\linewidth,height=0.3\linewidth]{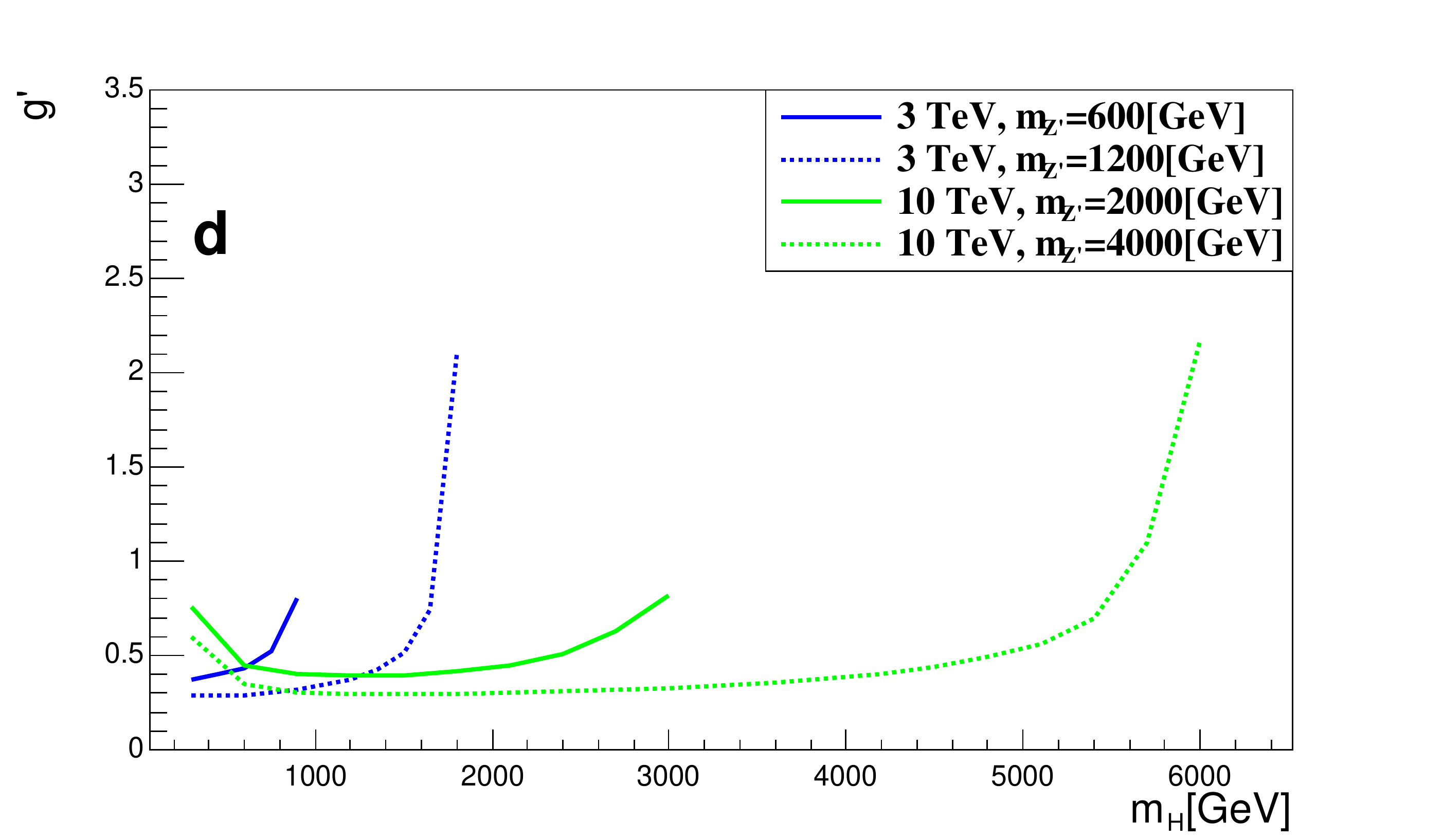}
	\end{center}
	\caption{  Panel (a): The $5 \sigma$ discovery region of the same-sign tetralepton signature in the $g'-m_{Z'}$ plane, where we fix $m_{Z'}=m_H=3m_N$. The red region is excluded by neutrino trident production at CCFR \cite{CCFR:1991lpl}. The orange and black lines are the projected projected sensitivity at HL-LHC \cite{delAguila:2014soa} and muon collider \cite{Huang:2021nkl}. Panel (b): The $5 \sigma$ discovery region of the same-sign tetralepton signature in the $m_N-m_H$ plane, where we fix $g'=0.6$. Panel (c): The $5 \sigma$ discovery reach of $g'$ as a function of $m_{Z'}/m_H$, where we fix $m_H=3m_N=900$ GeV. Panel (d): The $5 \sigma$ discovery reach of $g'$ as a function of $m_H$, where we fix $m_H=3m_N$.   The blue and green lines are the discovery limits at the 3~TeV and 10 TeV muon collider, respectively. 
		\label{fig5}}
\end{figure}

Based on the above analysis, we then investigate the promising region of the same-sign tetralepton signature. In the  panel (a) of Figure \ref{fig5}, we show the $5\sigma$ discovery region in the $g'-m_{Z'}$ plane, where we fix the relation $m_{Z'}=m_H=3 m_N$ for illustration. With growing production cross section, the discovery reach of $g'$ typically decreases as $m_N$ becomes larger before the kinematic threshold. Obviously, the 3 TeV muon collider is more promising to probe the region with $m_N\lesssim 500$ GeV. The 3 TeV muon collider could discover a large part of the regime with $g'\gtrsim0.38$ for the electroweak scale heavy neutral lepton. With higher integrated luminosity, the 10 TeV muon collider would unravel the parameter space with $g'\gtrsim0.4$.   Since we focus on the heavy $Z'$ above the electroweak scale, the only constraint is from neutrino trident production at CCFR \cite{CCFR:1991lpl}. The HL-LHC might cover the parameter space with $g'\gtrsim0.1$ and $m_{Z'}$ below the TeV scale. In the future, the direct search of new $L_\mu-L_\tau$ gauge boson $Z'$ at muon collider could probe $g'\gtrsim10^{-2}$ through the $\mu^+\mu^- \to \ell^+\ell^-(\ell=\mu,\tau)$ or $ \ell \bar{\ell}\gamma(\ell=\mu,\tau,\nu)$ channel \cite{Huang:2021nkl}, which would confirm the promising region of the same-sign tetralepton signature.

In the panel (b) of Figure \ref{fig5}, we depict the sensitive region in the $m_N- m_H$ plane for the scenarios of $m_H=0.5m_{Z'}$, $m_H=m_{Z'}$ and $m_H=2m_{Z'}$ with $g'=0.6$. For the benchmark scenario with $m_H=m_{Z'}$, we report that the 3 TeV muon collider could detect the parameter space with 400 GeV $\lesssim m_H\lesssim$ 1500~GeV, which covers a large part of the kinematically allowed region. The 10 TeV muon collider could probe the regime of 600 GeV $\lesssim m_H\lesssim$ 5000 GeV. We find that the 10 TeV muon collider is not promising to detect $m_N$ below about 150 GeV, which is mainly due to the non-isolation of the muon from the jet in the highly boosted heavy neutral lepton decays. Close to the threshold region with $m_{Z'}=m_H\sim 2 m_N$, the branching ratio of $H/Z'\to NN$ is also suppressed, so the same-sign tetralepton signature is not promising in such a region either.
For the benchmark scenario with $m_H=0.5m_{Z'}$, the upper limit of $m_H$ becomes  smaller than that of the scenario $m_H=m_{Z'}$. In contrast, the upper limit of $m_H$ becomes larger for the scenario with $m_H=2m_{Z'}$. This is because the upper limit of $m_H$ is mainly determined by the kinematic threshold $m_H+m_{Z'}<\sqrt{s}$. On the other hand, the upper limit of $m_N$ for $m_H=2m_{Z'}$ becomes smaller than that of $m_H=m_{Z'}$,  since it is roughly set by the threshold of $Z'\to NN$. 

In the panel (c) of Figure \ref{fig5}, we illustrate the impact of mass ratio $m_{Z'}/m_H$ on the discovery limit of $g'$, where $m_H=3m_N=900$ GeV is fixed. The lower limit of $g'$ becomes smaller as the mass ratio $m_{Z'}/m_H$ increases. When $m_{Z'}/m_H\lesssim1$, the branching ratio of $Z'\to NN$ is suppressed by the final state phase space, which leads to the quickly increasing of $g'$. As shown in panel (c) of Figure \ref{fig2}, the production cross section at the 10 TeV muon collider could be enhance by over 25 times compared with $m_{Z'}=m_H$. Therefore, the lower limit of $g'$ could decrease down to about 0.2 when $m_{Z'}/m_H\sim10$.  

In the panel (d) of Figure \ref{fig5}, we show the $5\sigma$ discovery region of $g'$ as a function of $m_H$ for some benchmark values of $m_{Z'}$. At the 3 TeV muon collider, the promising region is within $g'\gtrsim0.4$ and $m_H\lesssim900$~GeV when $m_{Z'}=600$ GeV. Notice that the upper limit of $m_H$ is determined by the threshold of $Z'\to NN$ as $m_H=3m_N<3m_{Z'}/2$, since we fix the mass relation $m_H=3m_N$.  The promising region extends to $g'\gtrsim0.3$ and $m_H\lesssim1800$ GeV when $m_{Z'}=1200$~GeV. Besides larger sensitive regions, the results are similar for the 10 TeV muon collider.

Although the same-sign tetralepton signature $4\mu^\pm+4J$ is the most exotic signal, it should not be the most promising one to discover the heavy neutral leptons $N$, new gauge boson $Z'$, and heavy Higgs $H$. For instance, when combining all the visible final states of $4 N$ without distinguishing the specific charge of muon, the cross section of the tetralepton signature $4\mu +4J$ is theoretically eight times larger than that of the same-sign one $4\mu^\pm +4J$. With a much larger cross section, the tetralepton signature might probe a larger parameter space by properly dealing with the backgrounds.

\section{Same-Sign Trilepton Signature}\label{SEC:SG2}

As shown in Figure \ref{fig1}, the leptonic $Z'\to\mu^+\mu^-$ is the dominant decay mode. This leptonic decay $Z'\to\mu^+\mu^-$ also  has much higher detection efficiency than the $Z'\to NN$ mode, so the same-sign trilepton signature is expected to be more promising than the same-sign tetralepton signature. In this scenario, lepton number violation is only from the heavy Higgs decay. The same-sign trilepton signature is generated via the process
\begin{equation}
	\mu^+\mu^-\to Z' H \to \mu^+\mu^- + NN \to \mu^+\mu^- + 2 \mu^\pm + 2 W^\mp \to   3\mu^\pm + \mu^\mp  +2J.
\end{equation}
We notice that the cross section of the background for the same-sign dilepton signature $2\mu^\pm +2J$ after certain selection cuts is at the order of $\mathcal{O}(10^{-3})$ fb \cite{He:2024dwh}. With two more muons in the final states, the cross section of the background for the same-sign trilepton signature $ 3\mu^\pm + \mu^\mp  +2J$ is roughly at the order of $\mathcal{O}(10^{-3})\times \alpha_\text{em}\sim10^{-5}$ fb. The resulting background event is less than one even with  10 ab$^{-1}$ data at the muon collider. For illustration, we also assume one background event $N_B=1$ in this section to calculate the significance. 

\begin{figure}
	\begin{center}
		\includegraphics[width=0.45\linewidth,height=0.3\linewidth]{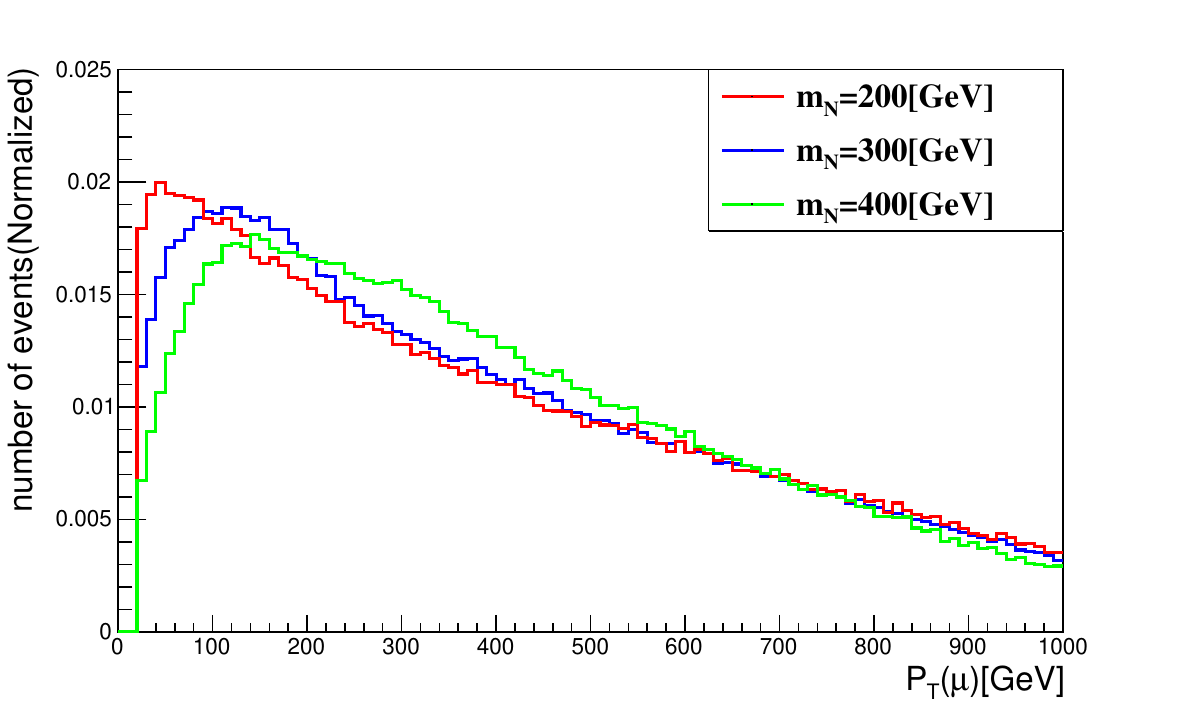}
		\includegraphics[width=0.45\linewidth,height=0.3\linewidth]{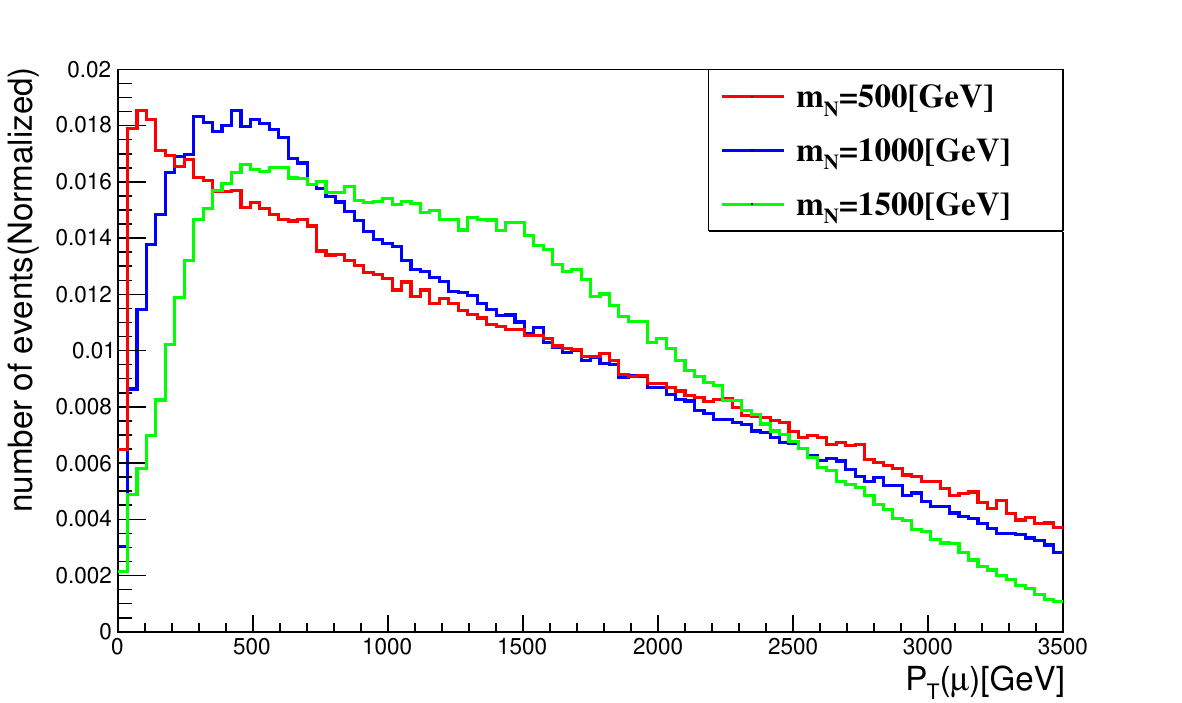}
		\includegraphics[width=0.45\linewidth,height=0.3\linewidth]{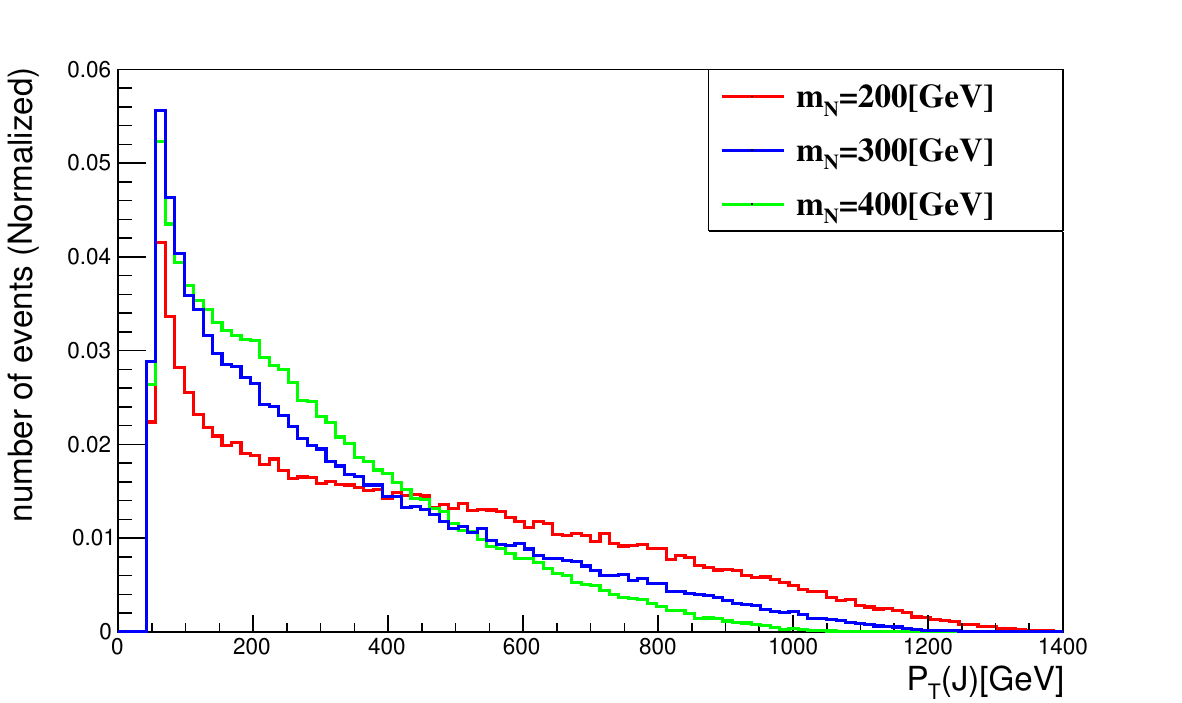}
		\includegraphics[width=0.45\linewidth,height=0.3\linewidth]{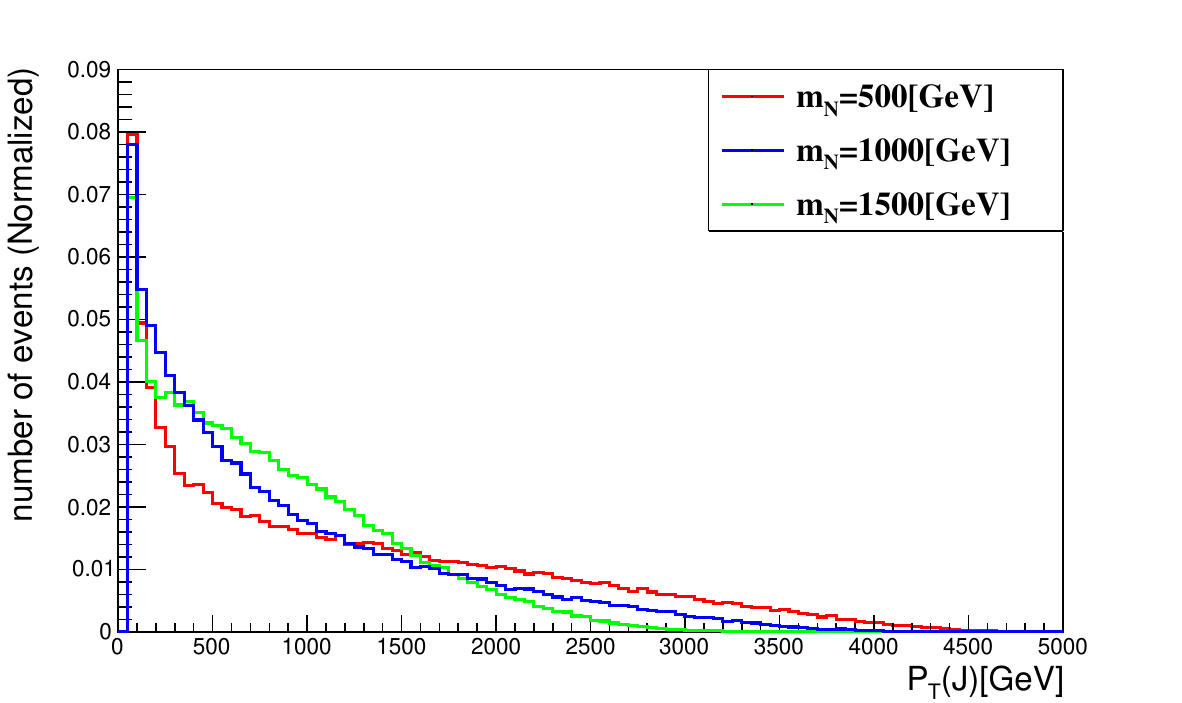}
		\includegraphics[width=0.45\linewidth,height=0.3\linewidth]{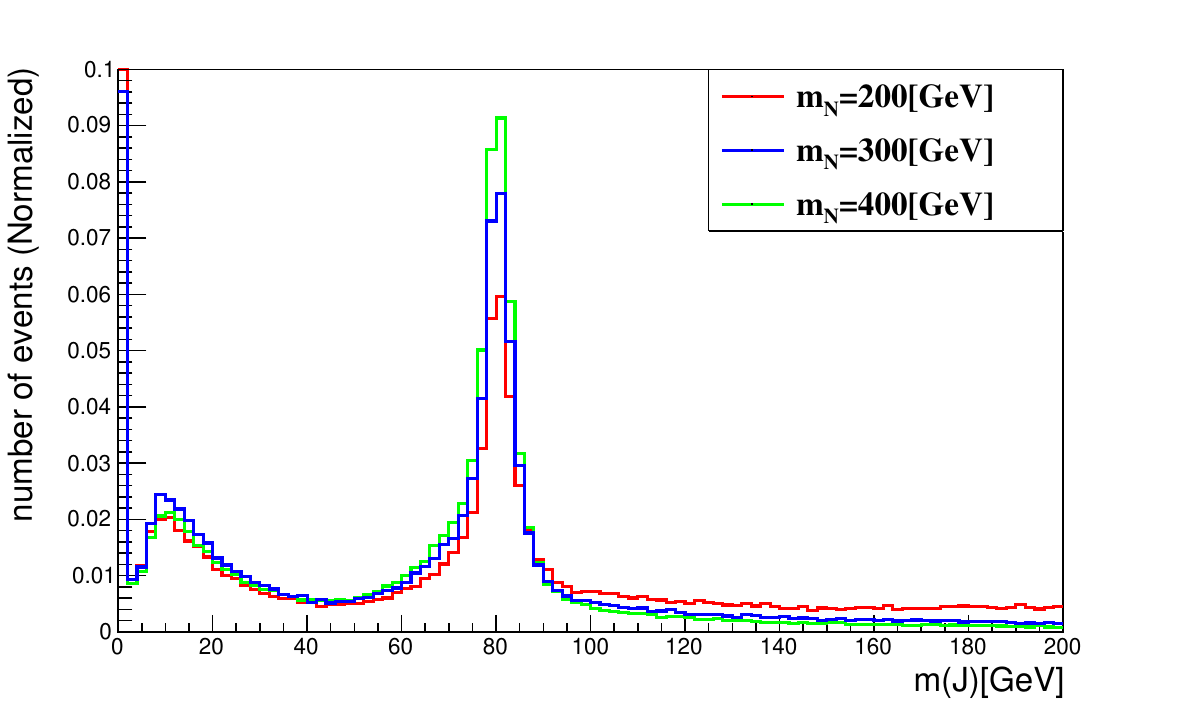}
		\includegraphics[width=0.45\linewidth,height=0.3\linewidth]{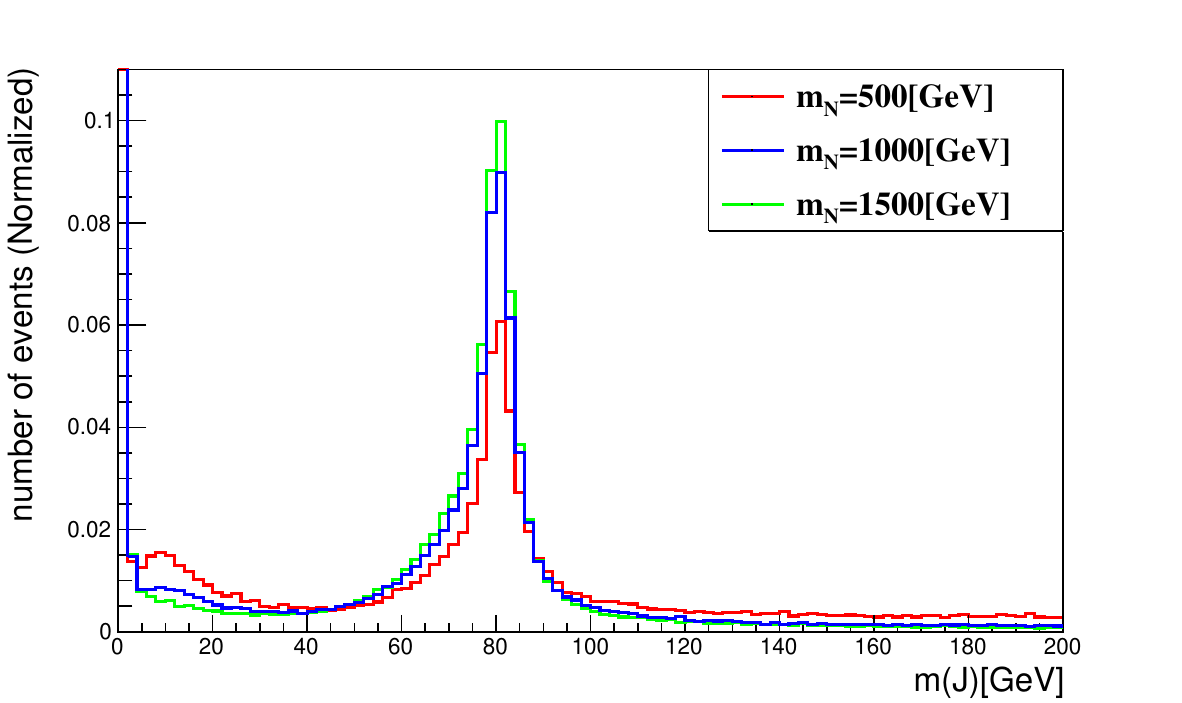}
	\end{center}
	\caption{Normalized distributions of muon transverse momentum $P_T(\mu)$, fat-jet transverse momentum $P_T(J)$, and invariant mass of fat-jet $m_J$. Left panels are the results of 3 TeV, and right panels are the results of 10 TeV. During the simulation of these plots, we have assumed the relation $m_{Z'}=m_H=3m_N$ for illustration.
		\label{fig6}}
\end{figure}

We employ the same pre-selection cuts in Equation \eqref{Eqn:ps} in this section. The same-sign trilepton signature is then extracted by the following selection criteria:
\begin{itemize}
	\item {\bf Lepton Selection: } $N(\mu^\pm)=3$, $N(\mu^\mp)=1$,
	\item {\bf Fat-jet Selection: } $N(J)=2$, 50 GeV$<m_J<$ 110 GeV.
\end{itemize}
All the two fat-jets in the final state are required to probe the heavy Higgs $H$.

In Figure \ref{fig6}, we show some variables of the same-sign trilepton signature after the pre-selection cuts. For the muon transverse momentum $P_T(\mu)$ of the same-sign trilepton signature, it tends to be larger than that of the same-sign tetralepton signature for the corresponding benchmarks. This is because there are two muons coming from the direct decay of gauge boson $Z'\to \mu^+\mu^-$ in the trilepton signature, while the cascade decay of heavy neutrino $N\to \mu^\pm J$ generates all muons in the tetralepton one. On the other hand, the fat-jet transverse momentum $P_T(J)$ and invariant mass $m_J$ of the same-sign trilepton signature are quite similar to those of the tetralepton signal, as they are all induced by the channel of $N\to \mu^\pm J$. We keep the same cut on $m_J$ to select fat-jet from $W$ decay.

\begin{table}
	\begin{center}
		\begin{tabular}{c | c | c | c | c |c} 
			\hline
			\hline
			\multicolumn{2}{c|}{$\mu^+\mu^-\to  3\mu^\pm + \mu^\mp+2J$} & Pre-Selection (fb) & After Selection (fb) & Significance & $5\sigma$ Luminosity (fb$^{-1}$)\\
			\hline
			\multirow{3}{*}{3 TeV}	& $m_N=200$ GeV & 3.20 & 1.73 & 149 & 4.82 \\			
			\cline{2-6}
			& $m_N=300$ GeV & 3.95 & 2.61 & 189 & 3.19 \\			
			\cline{2-6}
			& $m_N=400$ GeV & 4.56 & 3.03 & 206 & 2.75 \\			
			\hline
			\multirow{3}{*}{10 TeV}	& $m_N=500$ GeV & 0.251 & 0.129 & 126 & 64.6 \\			
			\cline{2-6}
			& $m_N=1000$ GeV & 0.353 & 0.229 & 175 & 36.4\\			
			\cline{2-6}
			& $m_N=1500$ GeV & 0.372 & 0.238 & 180 & 35.0 \\			
			\hline
			\hline
		\end{tabular}
	\end{center}
	\caption{Same as Table \ref{Tab01}, but for the same-sign trilepton signature $\mu^+\mu^-\to  3\mu^\pm + \mu^\mp +2J$.}
	\label{Tab02}
\end{table}

In Table \ref{Tab02}, we show the results of the same-sign trilepton signature at the muon collider, which is also under the assumption of $m_{Z'}=m_H=3 m_N$ and $g'=0.6$. The after selection cuts cross sections at the 3 TeV muon collider are 1.73 fb, 2.61 fb, and 3.03 fb for $m_N=$200 GeV, 300 GeV, and 400 GeV, respectively. With an integrated luminosity of 1 ab$^{-1}$, the significance could reach 149, 189, and 206, accordingly. All these 3 TeV benchmarks can be discovered with less than 5 fb$^{-1}$ data. For the three benchmarks at the 10 TeV muon collider, the typical cross section after selection cuts is at the order of $\mathcal{O}(0.1)$ fb . With 10 ab$^{-1}$ data, the significance of these 10 TeV benchmarks can also exceed 100. The benchmark with $m_N=$500~GeV, 1000 GeV, and 1500 GeV can be discovered at the 10 TeV muon collider with 64.6 fb$^{-1}$, 36.4 fb$^{-1}$, and 35.0~fb$^{-1}$ data. Compared to the same-sign tetralepton signal, the cross section of the same-sign trilepton signature is about two orders of magnitude larger for certain benchmarks. Therefore, this less exotic same-sign trilepton signature $ 3\mu^\pm + \mu^\mp  +2J$ is quite promising as the discovery channel of the heavy neutral lepton at the muon collider.

\begin{figure}
	\begin{center}
		\includegraphics[width=0.45\linewidth,height=0.3\linewidth]{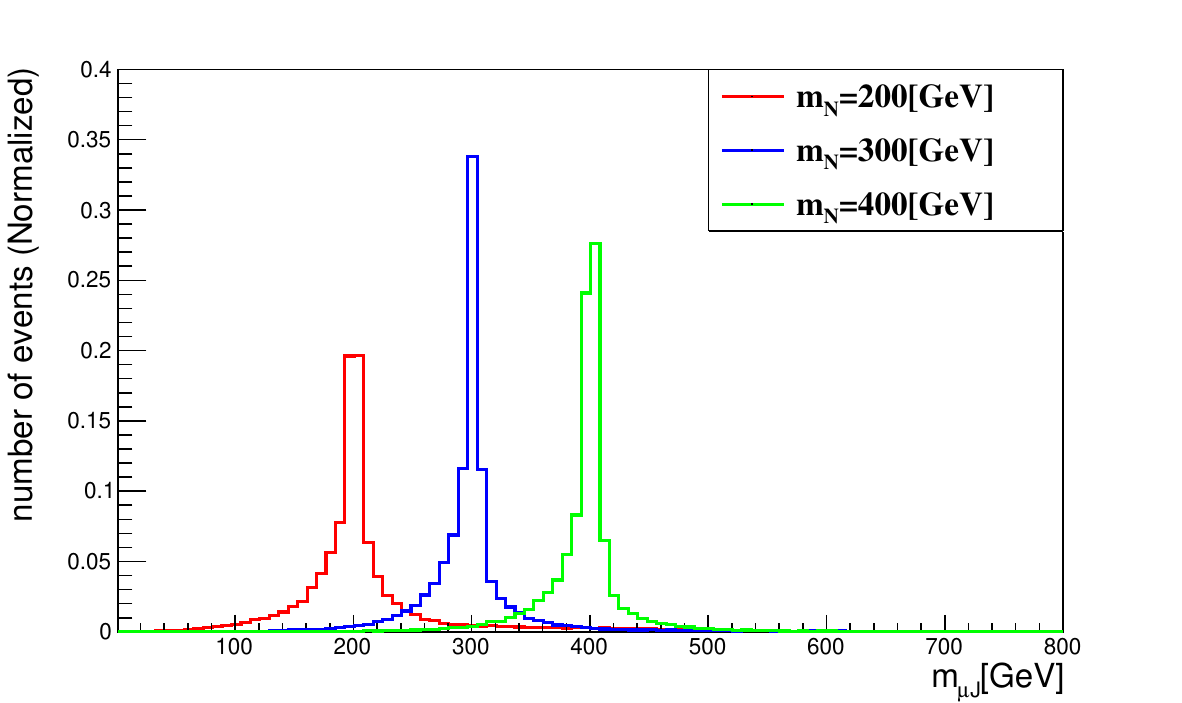}
		\includegraphics[width=0.45\linewidth,height=0.3\linewidth]{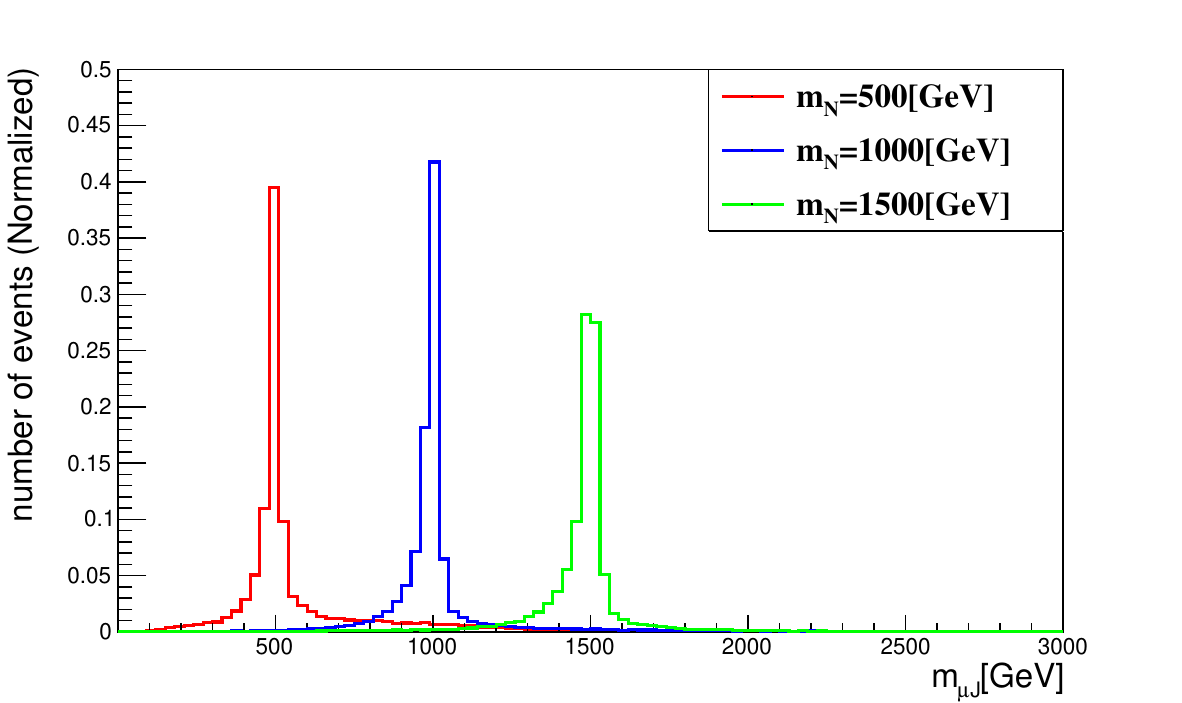}
		\includegraphics[width=0.45\linewidth,height=0.3\linewidth]{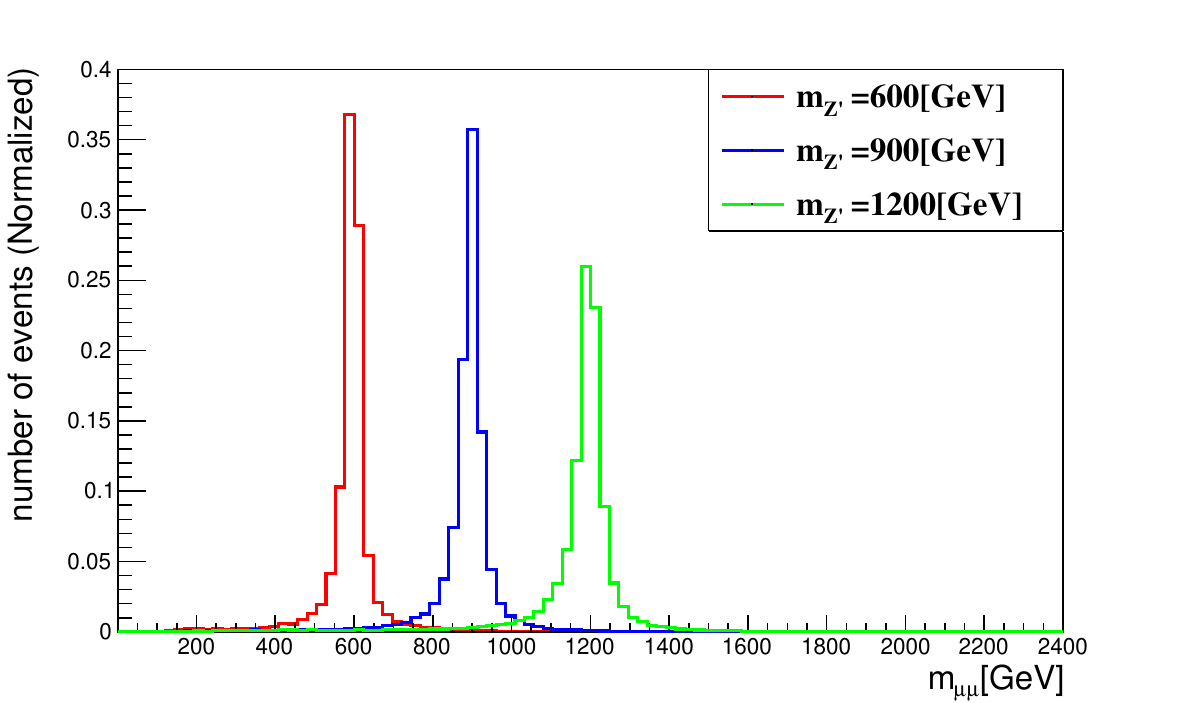}
		\includegraphics[width=0.45\linewidth,height=0.3\linewidth]{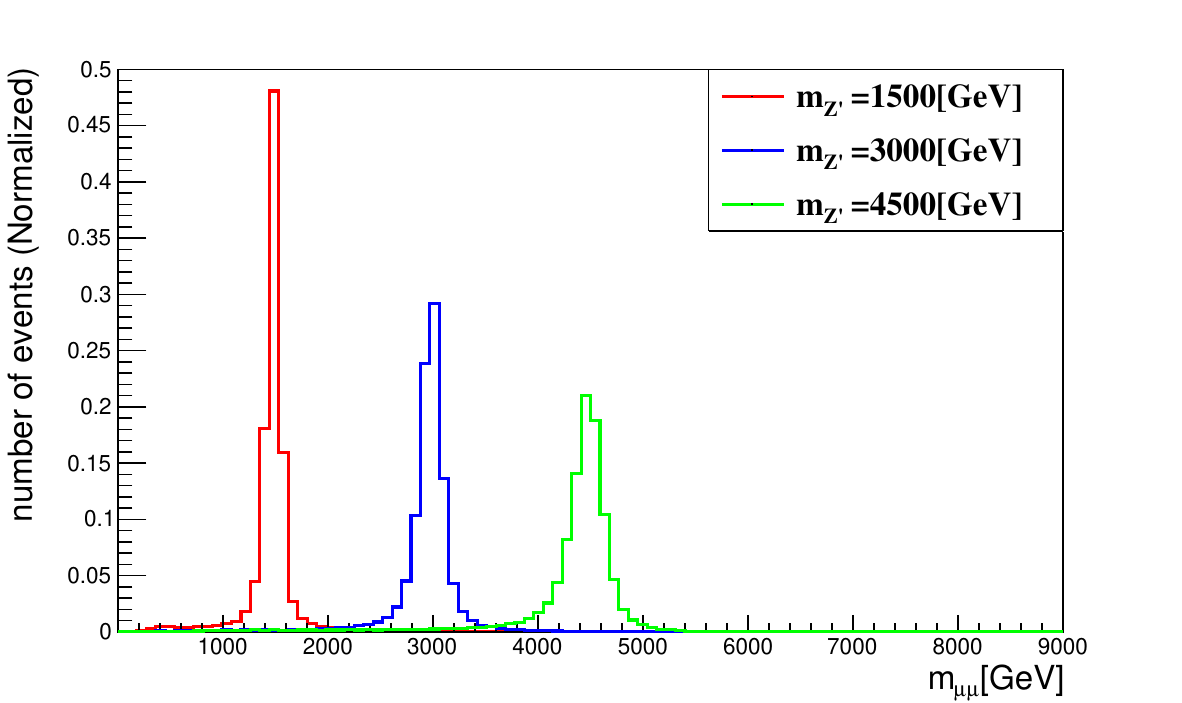}				
		\includegraphics[width=0.45\linewidth,height=0.3\linewidth]{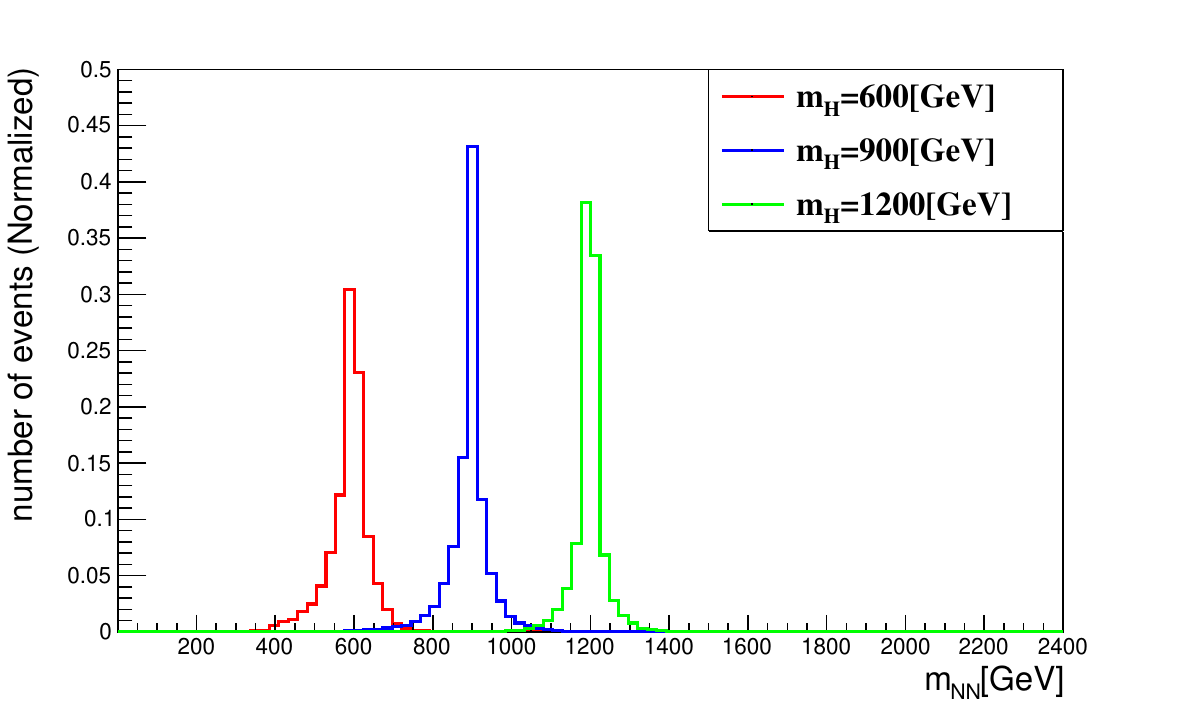}
		\includegraphics[width=0.45\linewidth,height=0.3\linewidth]{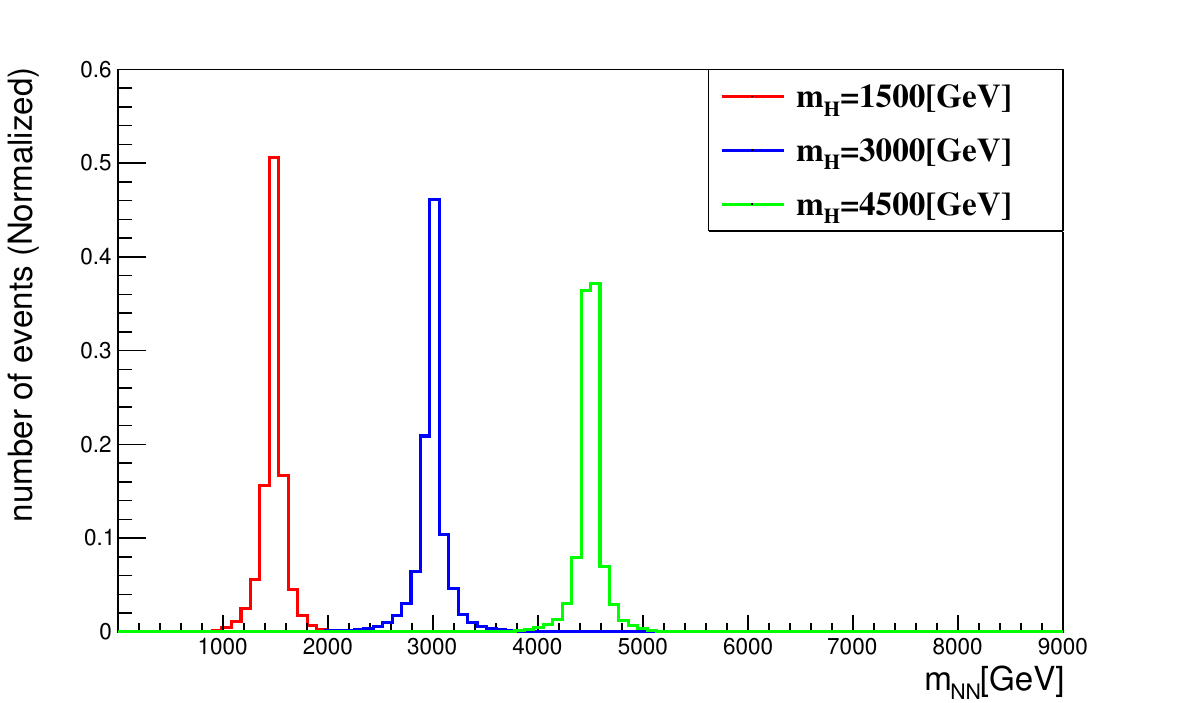}
		\includegraphics[width=0.45\linewidth,height=0.3\linewidth]{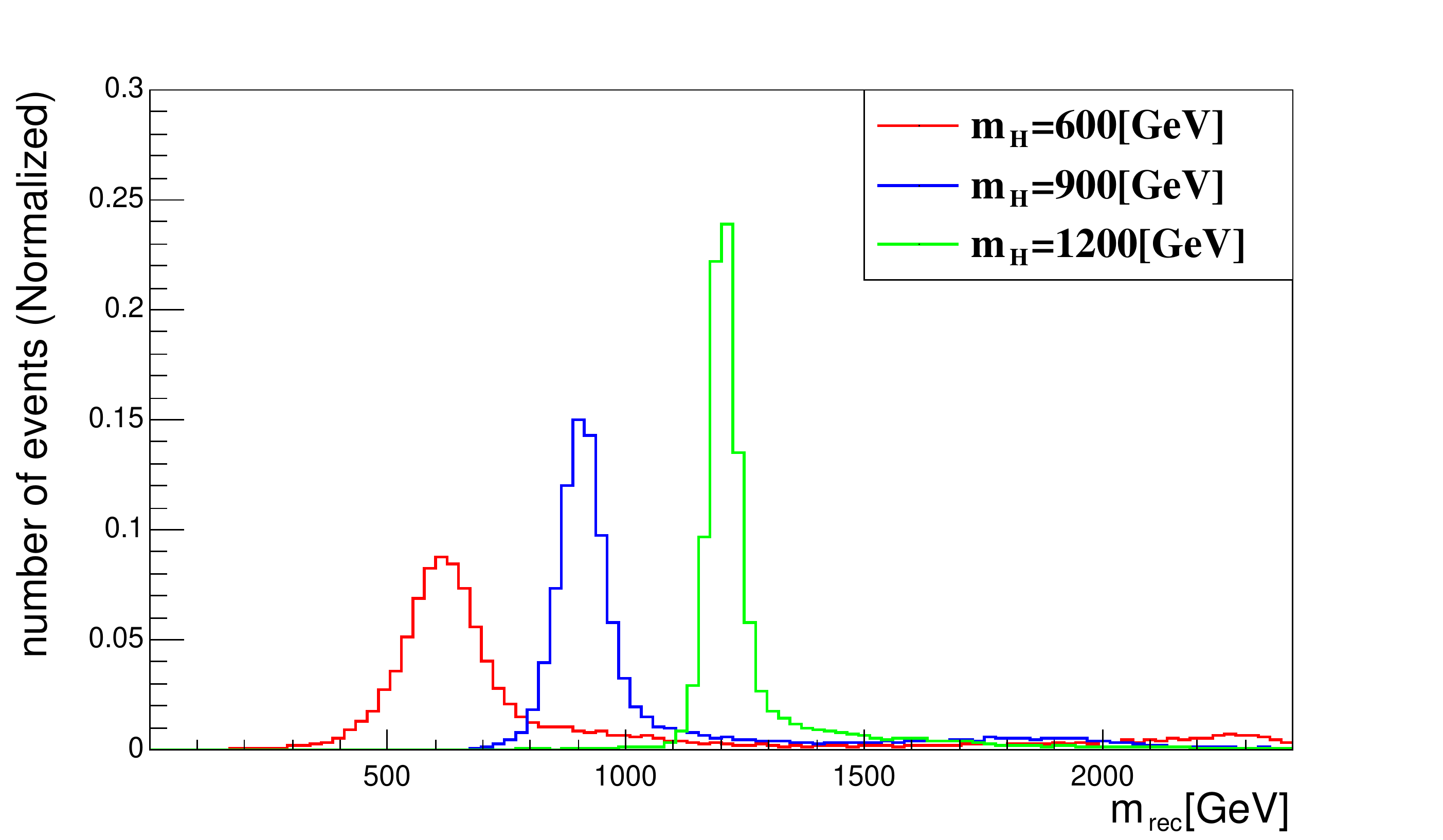}
		\includegraphics[width=0.45\linewidth,height=0.3\linewidth]{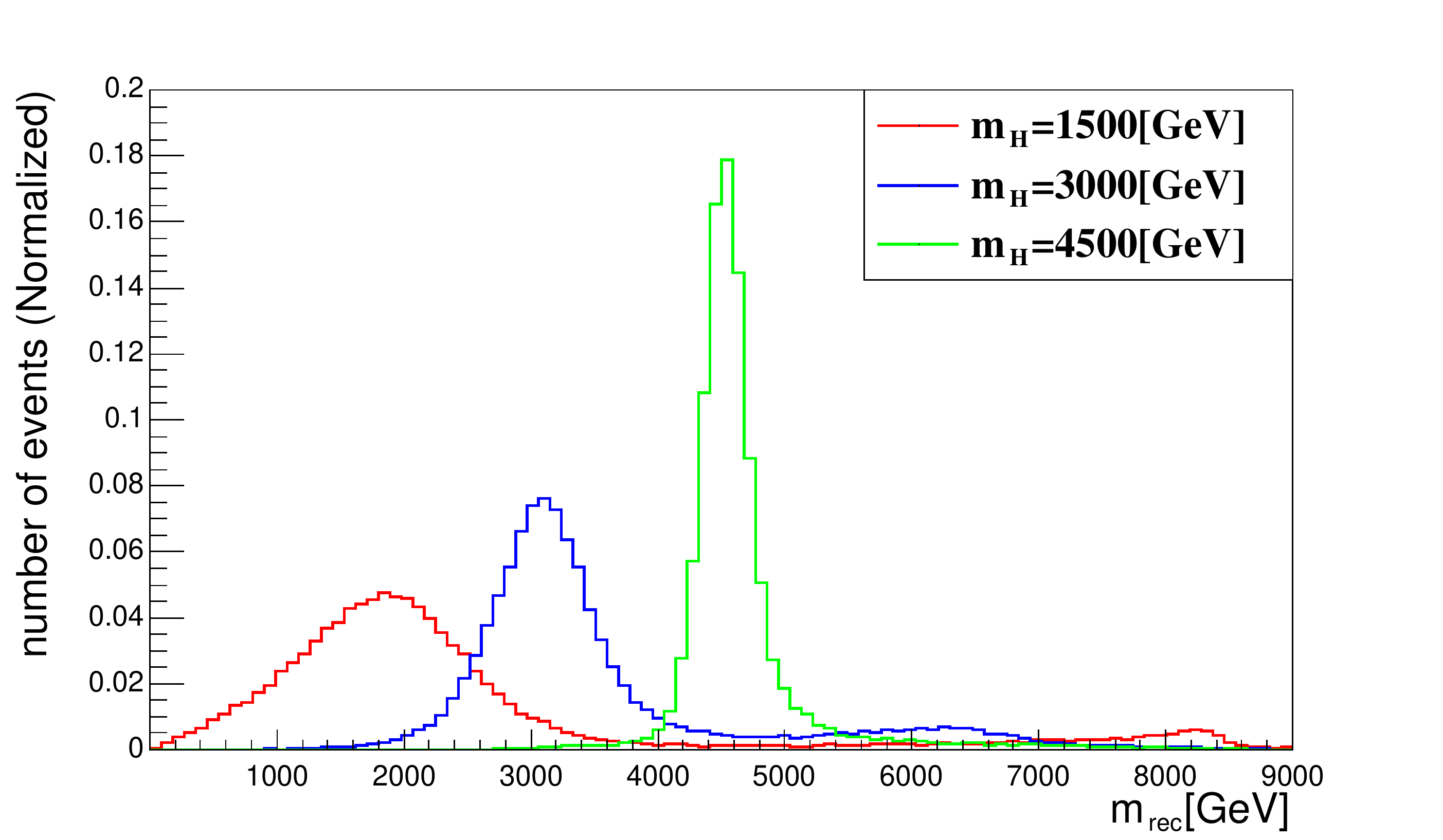}
	\end{center}
	\caption{Normalized distributions of invariant mass of muon and fat-jet $m_{\mu J}$, invariant mass of two opposite-sign muon $m_{\mu \mu}$, invariant mass of two reconstructed heavy neutral lepton $m_{NN}$, and recoil mass $m_\text{rec}$. Left panels are the results of 3 TeV, and right panels are the results of 10 TeV. The relation $m_{Z'}=m_H=3m_N$ is also assumed. 
		\label{fig7}}
\end{figure}

Within the same-sign trilepton signature, the three same-sign muons originate from different heavy resonances. One is from the direct decay $Z'\to \mu^+\mu^-$, and the other two are from the decay $N\to \mu^\pm J$. In this section, we perform a combined analysis by minimizing
\begin{equation}
	\chi^2_{NZ'}=\sum_{i,j,k}(m_{\mu^\pm_i J_j}-m_N)^2 +(m_{\mu^\pm_k \mu^\mp} - m_{Z'})^2,
\end{equation}
to reconstruct the heavy neutral lepton and new gauge boson simultaneously. Although we assume $m_{Z'}=m_H$ for illustration, this $\chi^2_{NZ'}$ function is applicable  for the scenario of $m_{Z'}\neq m_H$. The results are shown in the first two rows of Figure \ref{fig7}.

For the heavy Higgs, there are two viable pathways to reconstruct its mass. The first one is directly determined by the invariant mass of the two heavy neutral leptons $m_{NN}$, which originates from the decay mode $H\to NN$. Similar to the vanilla Higgs-strahlung process $\ell^+\ell^-\to Zh$, the second approach is considering the recoil mass of the visible decay $Z'\to \mu^+\mu^-$
\begin{equation}\label{Eqn:mrec}
	m^2_\text{rec}=(\sqrt{s}-E_{\mu^+\mu^-})^2-p^2_{\mu^+\mu^-}=s-2E_{\mu^+\mu^-}\sqrt{s}+m^2_{\mu^+\mu^-},
\end{equation}
where $E_{\mu^+\mu^-}$, $p_{\mu^+\mu^-}$, and $m_{\mu^+\mu^-}$ are the total energy, momentum and invariant mass of the opposite-sign muon pair $\mu^+\mu^-$ from $Z'$. The reconstructed heavy Higgs mass via invariant mass $m_{NN}$ and recoil mass $m_\text{rec}$ are shown in the third and fourth row of Figure \ref{fig7}. According to Equation \eqref{Eqn:mrec}, the resolution of recoil mass heavily depends on the energy resolution of the muon. Because of the smaller  energy detector resolution smearing at a lower muon energy \cite{Chakrabarty:2014pja}, we find that the spread of recoil mass peak decreases as the heavy Higgs mass increases. In this way, the direct invariant mass $m_{NN}$ is more precise than the recoil mass $m_\text{rec}$ to determine the heavy Higgs mass at the TeV scale muon collider.

\begin{figure}
	\begin{center}
		\includegraphics[width=0.45\linewidth,height=0.3\linewidth]{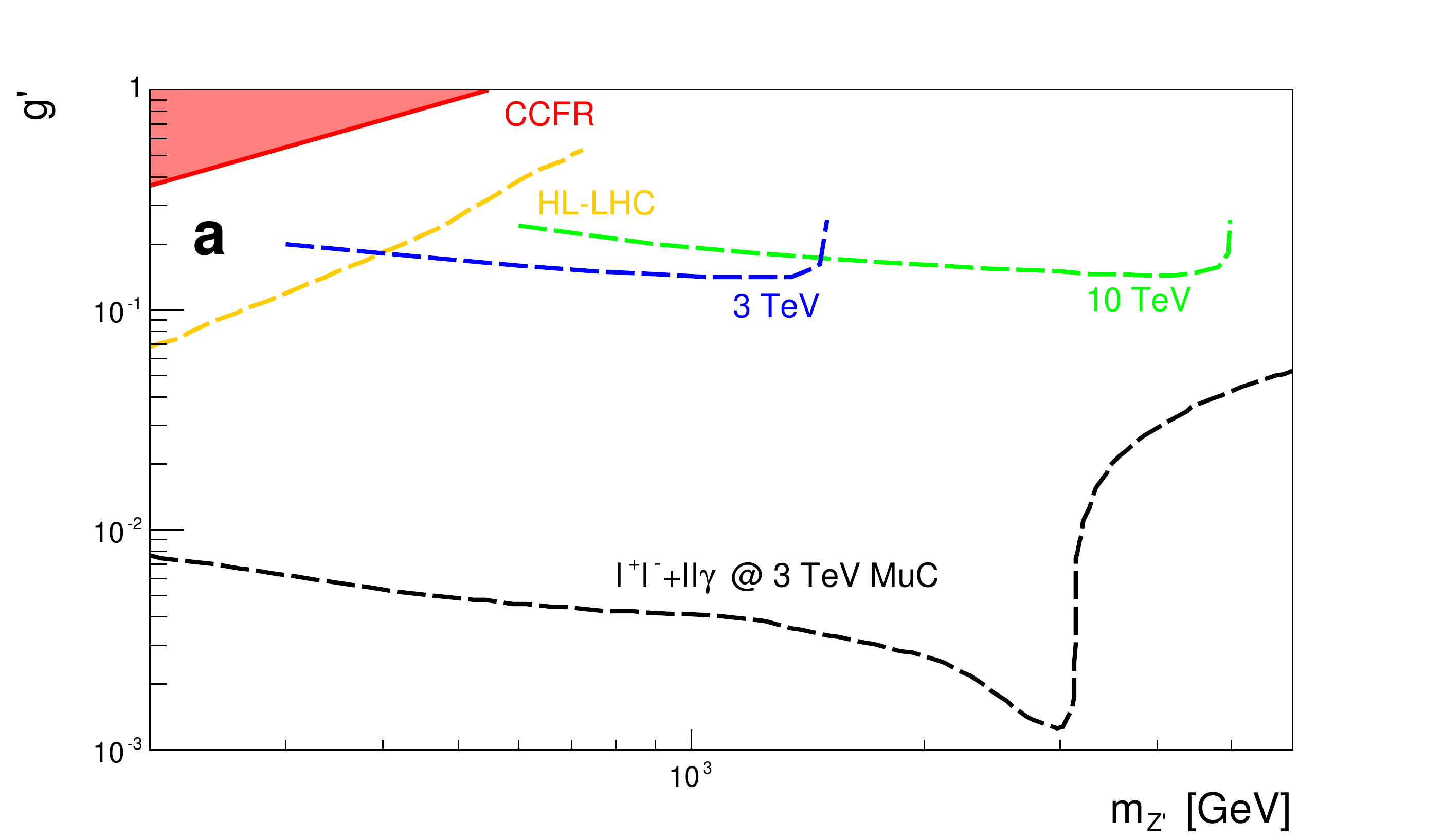}
		\includegraphics[width=0.45\linewidth,height=0.3\linewidth]{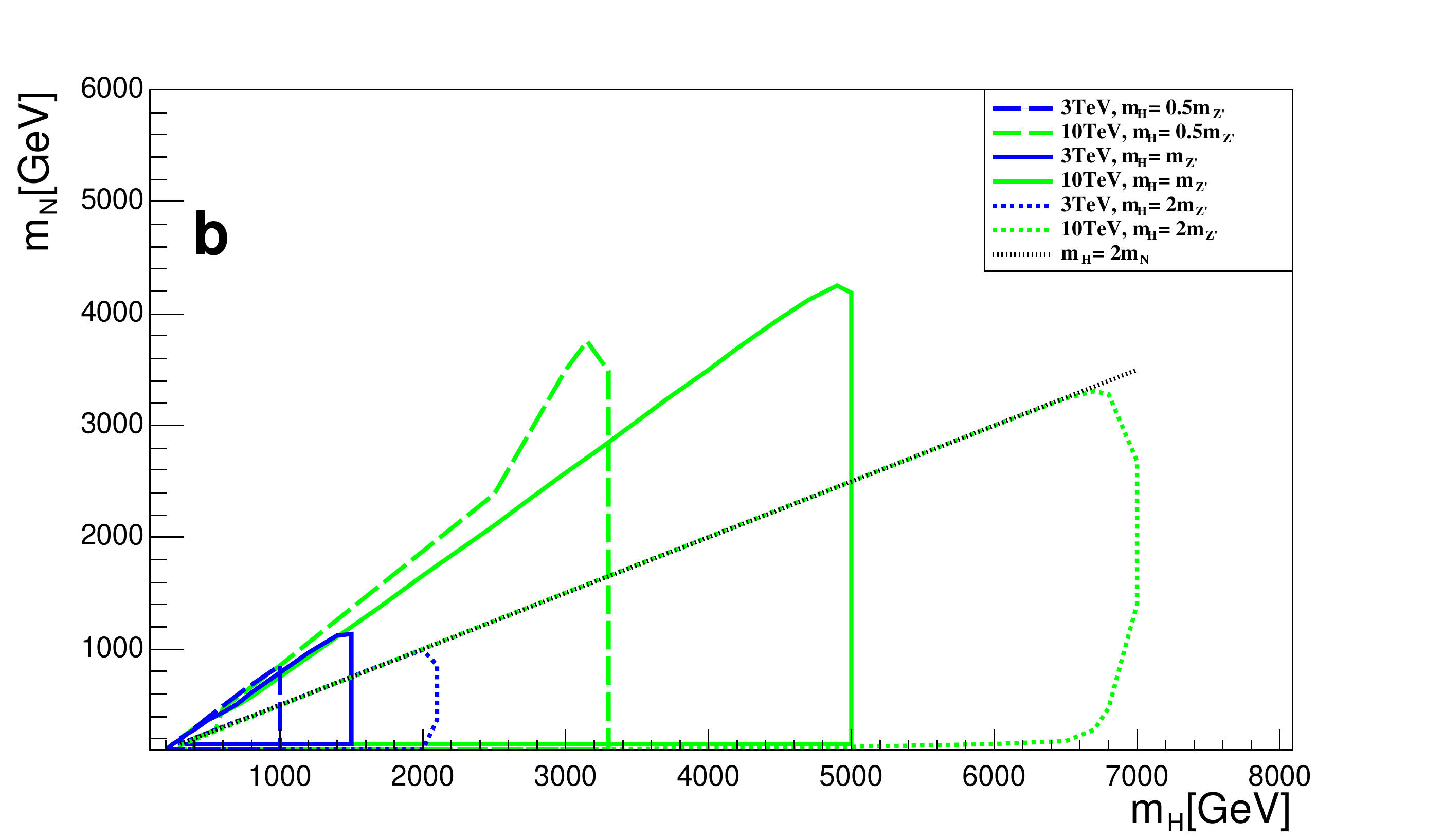}
		\includegraphics[width=0.45\linewidth,height=0.3\linewidth]{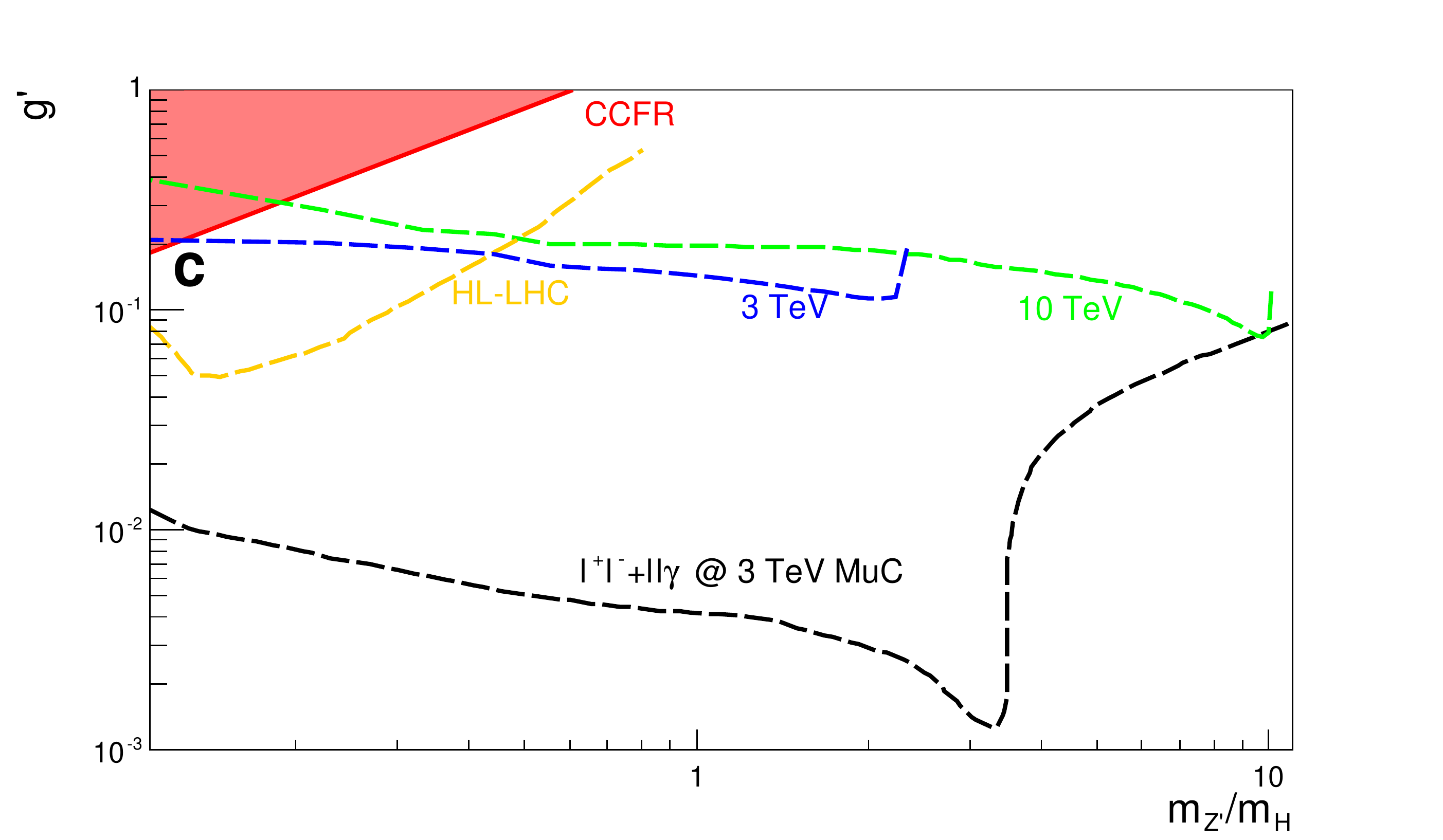}
		\includegraphics[width=0.45\linewidth,height=0.3\linewidth]{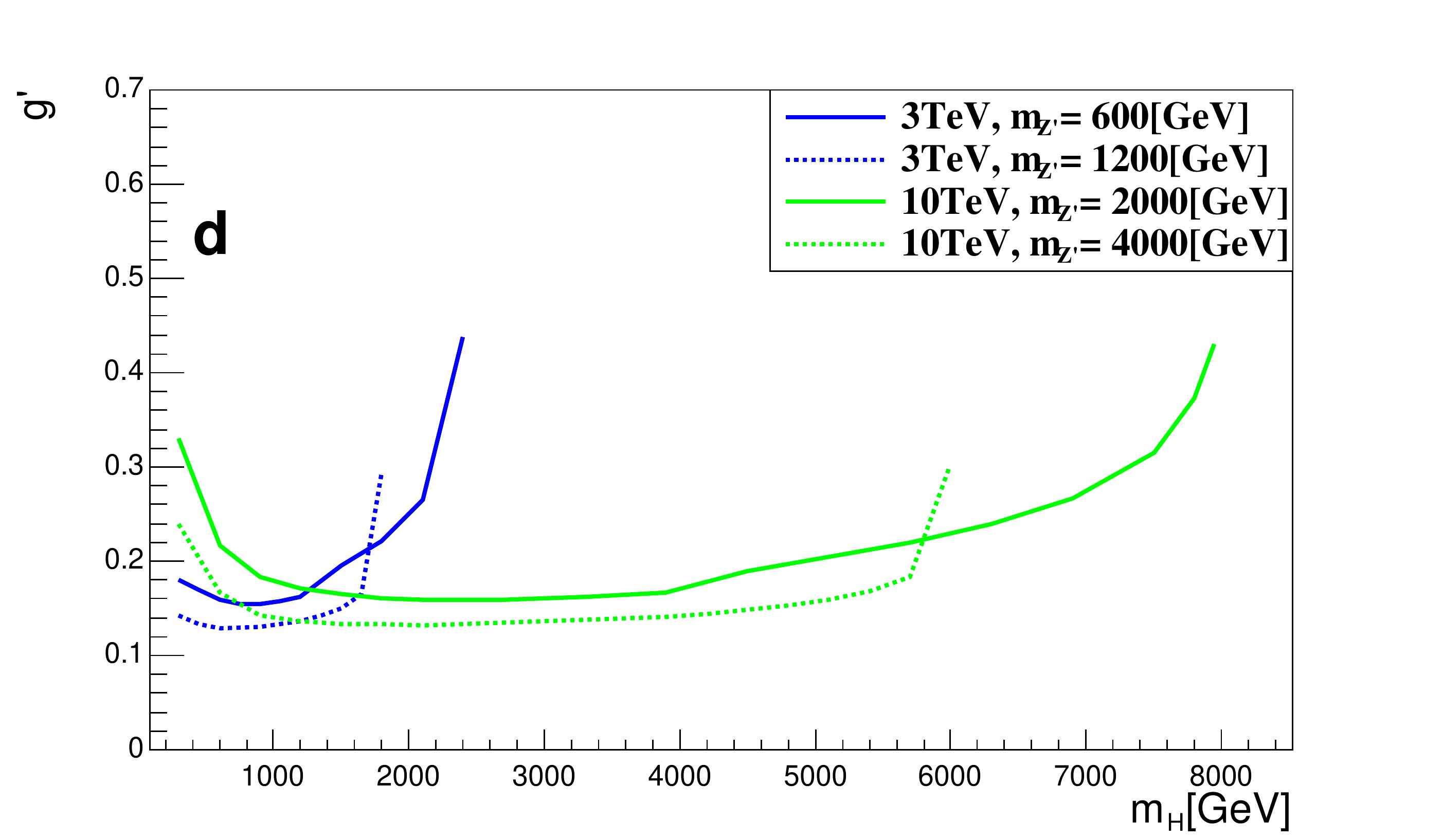}
	\end{center}
	\caption{  Panel (a): The $5 \sigma$ discovery region of the same-sign trilepton signature in the $g'-m_{Z'}$ plane, where we fix $m_{Z'}=m_H=3m_N$.   Panel (b): The $5 \sigma$ discovery region of the same-sign trilepton signature in the $m_N-m_H$ plane, where we fix $g'=0.6$. Panel (b): The $5 \sigma$ discovery reach of $g'$ as a function of $m_{Z'}/m_H$, where we fix $m_H=3m_N=900$ GeV. Panel (d): The $5 \sigma$ discovery reach of $g'$ as a function of $m_H$, where we fix $m_H=3m_N$.  The blue and green lines are the discovery limits at the 3~TeV and 10 TeV muon collider, respectively.  Other labels are the same as in Figure \ref{fig5}.  
		\label{fig8}}
\end{figure}

In the panel (a) of Figure \ref{fig8}, we depict the $5\sigma$ discovery region of the same-sign trilepton signature. The 3 TeV muon collider could discover the region with $g'\gtrsim 0.14$ for the electroweak scale heavy neutral lepton. Meanwhile, the 10 TeV is promising to discover heavy neutral lepton around the TeV scale when $g'\gtrsim0.15$. As shown in the panel (b) of Figure \ref{fig8},  we report that with a relatively large gauge coupling $g'=0.6$, the same-sign trilepton signature actually could probe the off-shell decay region when $m_H<2 m_N$  for the scenario with $m_H=0.5m_{Z'}$ and $m_{H}=m_{Z'}$. In such area, the lepton number violation Higgs decay is dominated by the three-body channel $H\to N^{*}N\to \mu^\pm W^\mp N \to \mu^\pm \mu^\pm JJ$ in the decoupling limit $\sin\alpha=0$. Above approximately the $m_N\gtrsim 0.8 m_H$ region, the four-body decay channel $H\to Z^{\prime *} Z^{\prime *}\to 4 \ell~(\ell=\mu,\tau)$ with $m_H=m_{Z'}$ becomes the dominant channel, which leads to the upper bounds of the same-sign trilepton signature.
For the scenario with a heavier $Z'$ as $m_H=0.5 m_{Z'}$, the contribution of the four-body channel is further suppressed. Therefore, a larger value of $m_N$ can be discovered for fixed value of $m_H$. We report that for the scenario with $m_H=2m_{Z'}$, the off-shell region $m_H<2m_N$ is not promising, because the four-body channel is dominant for such light $Z'$. As determined by the threshold $m_H+m_{Z'}<\sqrt{s}$, the upper limit of $m_H$ becomes smaller for $m_H=0.5 m_{Z'}$, while it becomes larger for $m_{H}=2m_{Z'}$.

In the panel (c) of Figure~\ref{fig8}, we show the discovery limit of $g'$ for the same-sign trilepton signature as function of $m_{Z'}/m_H$, where we fix $m_H=3m_N=900$ GeV. Increasing the mass ratio $m_{Z'}/m_H$ will also leads to the decreasing of $g'$. Different from the same-sign tetralepton signature, the branching ratio of $Z'\to \mu^+\mu^-$ is still allowed when $m_{Z'}<2 m_N$, so there are discovery region of the same-sign trilepton signature for  $m_{Z'}/m_H\ll 1$, which might be also tested by HL-LHC. Near the threshold $m_{Z'}+m_H\sim\sqrt{s}$, i.e., $m_{Z'}/m_H\sim10$ for the benchmark scenario at 10 TeV muon collider, the greatly enhanced cross section also leads to the clear decreasing of $g'$. 

In the panel (d) of Figure \ref{fig8}, we show the $5\sigma$ discovery region of $g'$ as a function of $m_H$ for some benchmark values of $m_{Z'}$. For TeV scale $m_H$, the muon collider could discover the region with $g'\gtrsim0.15$ through the same-sign trilepton signature. We report that increasing the gauge boson mass $m_{Z'}$ will decrease the upper limit of $m_H$ for muon collider with fixed energy, which actually corresponds to the threshold $m_{Z'}+m_H<\sqrt{s}$.

\section{Conclusion}\label{SEC:CL}

After the spontaneous symmetry breaking, we have the new gauge boson $Z'$ and heavy Higgs boson $H$ in the $U(1)$ gauged extension of the type-I seesaw mechanism. As the benchmark model, we consider the currently less constrained $U(1)_{L_\mu-L\tau}$ symmetry. These new bosons can be produced through the heavy Higgs-strahlung process $\mu^+\mu^-\to Z' H$ at the future muon collider, which typically reaches the maximum cross section near the threshold $\sqrt{s}\sim m_{Z'}+m_H$. In this way, the TeV scale muon collider is promising to detect these particles $Z'$ and $H$ also around the TeV scale. 

Further decays of $Z'$ and $H$, such as $Z'\to NN/\mu^+\mu^-,H\to NN$, lead to various novel signatures at the muon collider. In this paper, we investigate the same-sign tetralepton signature $\mu^+\mu^-\to Z' H\to NN+NN\to 4\mu^\pm+4J$ and the same-sign trilepton signature $\mu^+\mu^-\to Z' H\to\mu^+\mu^-+NN\to 3\mu^\pm + \mu^\mp +2J$ at the 3~TeV and 10 TeV muon collider. These lepton number violation signatures have a quite clean background. With a relatively larger production cross section, we find that the 3 TeV muon collider is usually more promising than the 10~TeV one to detect heavy neutral lepton around the electroweak scale. Meanwhile, the 10 TeV muon collider could discover TeV-scale heavy neutral lepton mainly due to larger luminosity.

The same-sign tetralepton signature $\mu^+\mu^-\to Z' H\to NN+NN\to 4\mu^\pm+4J$ is the most novel one, which violates the lepton number by four units. Based on the simplified scenario with $m_{Z'}=m_H$, we find that the muon collider could discover this novel signature when $g'\gtrsim 0.4$.  With a much larger cross section, the same-sign trilepton signature $\mu^+\mu^-\to Z'H\to \mu^+\mu^- + NN\to  3\mu^\pm + \mu^\mp  +2J$ is more promising as the discovery channel, which could probe the region with $g'\gtrsim0.15$. Limited by the decay $Z'\to NN$, the same-sign tetralepton signature is only viable within the regime of $m_{Z'}>2m_N$. On the other hand,  through the three-body decay $H\to N^{*}N\to \mu^\pm W^\mp N$ in the decoupling limit $\sin\alpha=0$, the same-sign trilepton signature could also test the off-shell decay region when $m_H<2 m_N$.

The most general scenario with $m_{Z'}\neq m_H$ can be easily confirmed by comparing the invariant mass $m_{\mu^+\mu^-}$ and recoil mass $m_\text{rec}$ of the muon pair from $Z'\to \mu^+\mu^-$.  Varying the mass of gauge boson $Z'$ would also alert the promising parameter space of these novel signatures. For instance, the discovery region is roughly limited by the threshold $m_{H}+m_{Z'}<\sqrt{s}$. Therefore, increasing the mass of $Z'$ will decrease the upper limit of $m_H$. On the other hand, a larger $m_{Z'}$ leads to a larger signal cross section, thus a lower limit of $g'$. For the same-sign tetralepton signature $\mu^+\mu^-\to Z' H\to NN+NN\to 4\mu^\pm+4J$, the upper limit of $m_N$ is determined by the lighter one of $Z'$ and $H$. For the same-sign trilepton signature $\mu^+\mu^-\to Z'H\to \mu^+\mu^- + NN\to  3\mu^\pm + \mu^\mp  +2J$, the off-shell region with $m_H<2 m_N$ is promising when $Z'$ is heavier than $H$. Meanwhile, the off-shell region becomes unpromising for relatively light $Z'$, such as $m_H\geq m_{Z'}$. 

In principle, our results are also applicable to the $U(1)_{B-L}$ symmetry. However, under the stringent constraints from dilepton search $Z'\to \ell^+\ell^-$, the promising regions of these novel signatures at the 3 TeV muon collider are already excluded by the current limit $m_{Z'}>5.1$ TeV. At the 10 TeV muon collider, there is still a viable region with $m_H<m_{Z'}$ and $m_H+m_{Z'}<10$ TeV. The larger parameter space of $U(1)_{B-L}$ can be tested at a muon collider with higher collision energy, such as the 30 TeV option.

\section*{Acknowledgments}

This work is supported by the National Natural Science Foundation of China under Grant No. 12505112, Natural Science Foundation of Shandong Province under Grant No. ZR2022MA056 and ZR2024QA138, State Key Laboratory of Dark Matter Physics, and University of Jinan Disciplinary Cross-Convergence Construction Project 2024 (XKJC-202404).


\end{document}